\documentclass[12pt,letterpaper]{article}

\usepackage{amsmath}
\usepackage{amsfonts}
\usepackage{amssymb}
\usepackage{amsthm}
\usepackage{bbm}
\usepackage{graphicx}
\usepackage{subcaption}
\usepackage{mathrsfs}
\usepackage{xcolor}
\usepackage{hyperref}
\usepackage{url}
\usepackage{verbatim}
\usepackage{titletoc}
\usepackage{color}
\usepackage{graphicx}
\usepackage{subcaption}
\usepackage{pdflscape}
\usepackage{comment}

\usepackage[letterpaper]{geometry}
\usepackage{setspace}
\allowdisplaybreaks

\usepackage[authoryear]{natbib}

\usepackage[subtle]{savetrees}
\usepackage[small,compact]{titlesec}

\newtheoremstyle{break}
  {\topsep}{\topsep}%
  {\itshape}{}%
  {\bfseries}{}%
  {\newline}{}%
\theoremstyle{break}
\newtheorem{theorem}{Theorem}

\newtheorem{proposition}{Proposition}

\newtheorem{lemma}{Lemma}

\usepackage[font=small, labelfont=bf]{caption}

\DeclareMathOperator*{\argmax}{arg\,max}
\DeclareMathOperator*{\argmin}{arg\,min}

\begin{document}

\title{Numerical Analysis of Test Optimality\footnote{Ketz gratefully acknowledges support from the Agence nationale de la recherche under Grant ANR-25-CE26-4320. McCloskey gratefully acknowledges support from the National Science Foundation under Grant SES-2341730.}
}
\author{Philipp Ketz\footnote{Paris School of Economics - CNRS, philipp.ketz@psemail.eu} \qquad Adam McCloskey\footnote{Department of Economics, University of Colorado, Boulder, adam.mccloskey@colorado.edu} \qquad Jan Scherer\footnote{Institute of Finance and Statistics, Department of Economics, University of Bonn, jscherer@uni-bonn.de}  
}

\maketitle

\begin{abstract}
In nonstandard testing environments, researchers often derive \emph{ad hoc} tests with correct (asymptotic) size, but their optimality properties are typically unknown a priori and difficult to assess. This paper develops a numerical framework for determining whether an ad hoc test is \emph{effectively} optimal—approximately maximizing a weighted average power criterion for some weights over the alternative and attaining a power envelope generated by a single weighted average power–maximizing test. Our approach uses nested optimization algorithms to approximate the weight function that makes an ad hoc test’s weighted average power as close as possible to that of a true weighted average power–maximizing test, and we show the surprising result that the rejection probabilities corresponding to the latter form an approximate power envelope for the former. We provide convergence guarantees, discuss practical implementation and apply the method to the weak-instrument–robust conditional likelihood ratio test and a recently-proposed test for when a nuisance parameter may be on or near its boundary.

\bigskip


\bigskip

\end{abstract}

\thispagestyle{empty}
\setcounter{page}{0}

\newpage

\section{Introduction}\label{sec:intro}

In nonstandard testing environments involving composite null and/or alternative hypotheses, a uniformly most powerful test often does not exist.  It is therefore common for researchers to devise implementable \emph{ad hoc} tests that are designed to have provably correct (asymptotic) size.  Although researchers may go to great lengths to demonstrate that their devised ad hoc tests have ``good'' power properties via Monte Carlo simulation, the theoretical power optimality properties of these tests are frequently unknown and elusive.  It is therefore difficult for both those who design hypothesis tests and those who are meant to use them to assess whether the power of a given test can be improved upon by a different test, a central question for its usefulness.

Instances for which researchers have sought to show that an ad hoc test is theoretically optimal in some sense have been limited.  Such approaches can be roughly divided into two types of power comparisons.  The first compares the power function of the ad hoc test to a point-wise power envelope. This point-wise power envelope is constructed as the power of a \emph{collection} of point-optimal tests computed at each point in the corresponding \emph{collection} of point alternative hypotheses for points lying in the composite alternative space (e.g.,~\citealp{AMS06}; \citealp{AMY19} and \citealp{GKM19}). Although it is possible for the power function of an ad hoc test to coincide with a power envelope constructed in this fashion, this is rarely the case because this form of power envelope may be unattainable by the power of any \emph{single} test.  To avoid or at least mitigate such unfair comparisons, researchers typically impose additional constraints on the point-optimal tests that are used to produce the point-wise power envelope such as unbiasedness (e.g.,~the two-sided $t$-test and \citealp{MM13,MM19}), rotational invariance (e.g.,~\citealp{AMS06} and \citealp{MSR21}) and similarity (e.g.,~\citealp{AMS06} and \citealp{MO20}).  However, the imposition of constraints on the underlying collection of (constrained) point-optimal tests still does not generally produce a power envelope that is necessarily attainable by a single test and leads to an optimality analysis that is limited to tests satisfying the constraints.  So when the power function of an ad hoc test falls short of a power envelope constructed in this way, what is a researcher to conclude?  Is the ad hoc test suboptimal?  Or is the power envelope simply an unfair and unattainable comparison because it does not correspond to the power function of any single test of the composite hypothesis?

The second approach to test optimality assessment in the literature compares either (i) the power function of the ad hoc test to the power function of a test that is known to be weighted average power (WAP) maximizing for a particular weight function over the parameter values in the composite alternative or (ii) the WAPs of these two tests, or both (e.g.,~\citealp{AP94}; \citealp{AMS08}; \citealp{EM14}; \citealp{EMW15} and \citealp{Cox24}).  This approach is also incomplete, leading to its own set of questions.  How should the researcher choose the weight function when constructing the WAP-maximizing test under comparison?  If the power function or WAP of the ad hoc test is not dominated by that of the WAP-maximizing test, should we conclude that it is optimal or could it be dominated by a WAP-maximizing test for a different set of weights?

In this paper, we propose to numerically determine whether there exists a weight function over the composite alternative that justifies the ad hoc test.  More precisely, we propose a numerical procedure for automatically finding the weight function that minimizes the WAP difference between the WAP-maximizing test corresponding to that weight function and the ad hoc test over a discretization of the composite alternative space. This eliminates the ambiguity of the second approach to test optimality implicit in the questions above.  When this WAP-difference is (not) approximately equal to zero, this approach immediately allows us to conclude that the ad hoc test under study is (not) effectively optimal in the sense of (not) being an approximately WAP-maximizing test itself.  A side benefit to our approach is that in the case for which the ad hoc test is determined to be effectively optimal, we know which weights over the alternative space make it approximately WAP-maximizing, allowing us to interpret how the test ``directs power''.

In addition to determining whether an ad hoc test is approximately WAP-maximizing for some weight function, we show the surprising result that our procedure immediately produces an attainable power envelope for the ad hoc test.  Over a discretization of the composite alternative space, we provide a set of sufficient conditions under which the power function of the WAP-maximizing test using the weights that minimize the WAP difference with the ad hoc test is a power envelope for the ad hoc test.   Since this power envelope is constructed from a single test, it is indeed attainable.  In addition, this power envelope is ``most favorable'' to the ad hoc test in the sense that it is constructed from a single WAP-maximizing test with WAP as close as possible to that of the ad hoc test.

Our numerical procedure is composed of an inner and an outer loop algorithm.  The inner loop algorithm computes an approximate WAP-maximizing test for a given set of weights over the discretized alternative space, in analogy with \cite{MM13}, \cite{EMW15} and \cite{AFBMOQST25}.  Our outer loop algorithm, which has no counterparts that we are aware of in the literature, approximates the weight function that minimizes the WAP difference between the corresponding WAP-maximizing test and the ad hoc test over the discretized alternative space.\footnote{\cite{KM25} made the first attempt we are aware of at numerically finding weights that minimize the WAP difference between an ad hoc test and a WAP-maximizing test.  However, their approach was confined to the particular problem of testing with inequality restrictions on nuisance parameters, carried no formal convergence guarantees and did not contain any results on producing power envelopes.  Nevertheless, it helped motivate us to write the current paper.}  Using the results mentioned above, the power function of this corresponding WAP-maximizing test can then be used as a power envelope for the ad hoc test.  We provide new convergence results for both the inner and outer loop algorithms, formally guaranteeing that they approximate the WAP-maximizing test and weights mentioned above.  Our algorithms and their convergence guarantees are general and not case-specific.

One may wonder why it would be more desirable to construct an ad hoc test with correct size and subsequently analyze its optimality properties rather than simply constructing a WAP-maximzing test directly using, e.g.,~\cite{MM13} or \cite{EMW15}.  There are two main reasons:~one conceptual and one computational.  The first reason for constructing an ad hoc test is that it is often unclear which weight function one should use when constructing a WAP-maximizing test.  Our procedure instead simply tells the researcher whether there exists a weight function that rationalizes the ad hoc test they have devised (as well as providing an approximation to it).  The second reason is that even with a clear idea of a desirable weight function, computing the corresponding WAP-maximizing test can become computationally prohibitive when the dimension of the alternative parameter space is too high.  Since our inner loop algorithm also computes WAP-maximizing tests iteratively, it is also of course subject to this criticism.  However, our numerical procedure can be used to provide partial evidence on the optimality of a test if the researcher applies it to lower-dimensional special cases and is able to show that the ad hoc test is optimal in those special cases.  On the other hand, a WAP-maximizing test for a lower-dimensional special case does not readily generalize up to higher dimensions.  

Finally, we note that discovering an ad hoc test to be effectively optimal for a particular testing problem can be used to motivate the development of tests for related problems.  For example, the theoretical optimality results established by \cite{AMS06,AMS08} in the specific setting of a homoskedastic linear IV model with randomly sampled data for \citeauthor{Mor03}'s (\citeyear{Mor03}) conditional likelihood ratio (CLR) test, apparently helped motivate \cite{AM16} and \cite{AG19} to produce generalizations of this test to settings that allow for dependent data, nonlinear models and singular variance matrices.

We apply our results and algorithms to two different testing problems, the first of which has been previously analyzed using the first approach to optimality assessment described above and the second of which has been analyzed using the second.  For the first problem, we substantively contribute to the ongoing debate over the optimality properties of the CLR test.  Specifically, we revisit classic optimality results from \cite{AMS06,AMS08} and more recent critiques by \cite{AMY19} and \cite{VdSW23} in the context of the Gaussian homoskedastic linear IV model.  Using the first existing approach to test optimality assessment described above, \cite{AMS06,AMS08} show that the CLR test effectively attains the point-wise power envelope for invariant (similar) tests under a design that holds the reduced-form error variance matrix constant.  On the other hand, \cite{AMY19} argue that the CLR test falls short of the point-wise power envelope obtained by fixing the true value of the parameter of interest and instead varying its hypothesized value, which \cite{VdSW23} show is equivalent to producing the more standard point-wise power envelope that fixes the hypothesized value and varies the true value under a design that instead holds the variance matrix of the structural errors constant.  Using our new algorithms, we show that the CLR test's power function is virtually indistinguishable from that of a \emph{single} WAP-maximzing test \emph{for both designs}, leading us to conclude that the CLR test is in fact effectively optimal in the class of invariant asymptotically efficient tests for this problem.

For our second application, we analyze the optimality properties of the new inequality-imposed confidence interval (IICI) of \cite{Cox24}, designed to improve inference when a scalar nuisance parameter may lie on or near the boundary of the parameter space.  Using the second existing approach to test optimality assessment described above, \cite{Cox24} compares the WAP of the test implied by the IICI to that of the WAP-maximizing test proposed for this problem by \cite{EMW15}, finding them to be very close.  We plot the power functions of both tests and find that they intersect, making the optimality properties of the IICI-implied test unclear.  Instead, we show that while the power function of the IICI-implied test is very close to the power function of a \emph{single} most favorable WAP-maximizing test, it does fall slightly short with a maximum gap of about 0.3 percentage points.  Nevertheless, its WAP is extremely close to that of the WAP-maximizing test, implying that it is highly-competitive but slightly short of optimal for problems with a scalar nuisance parameter that may lie on or near its boundary.

The scope for other applications of our results and numerical procedure is very wide and we just briefly mention some additional examples here.  In addition to analyzing the CLR test, one may seek to analyze the (unknown) optimality properties of other weak IV-robust tests in the literature such as the Lagrange multiplier test of \cite{Kle02}, which could be particularly interesting due to its non-monotonic power function.  \citeauthor{CY06}'s (\citeyear{CY06}) test of stock return predictability was not found to be dominated by the WAP-maximizing test with weights chosen by \cite{EMW15}.  Using our approach, it would be interesting to learn if there is a WAP-maximizing test that indeed dominates it.  We could use our approach to determine whether the apparent power deficiencies of the test of \cite{GKM19} when compared to the point-wise power envelope disappear when using our approach that makes the attainable comparison to a most favorable single WAP-maximizing test.  More broadly, our approach could be used to analyze tests in the large literatures on inference for moment inequality models, inference robust to identification failure, inference in structural change models and inference with highly-persistent (e.g.,~local-to-unit root) processes.  We intend to analyze some of these examples in follow-up work.  And we of course hope that our results and numerical procedures will be useful for the optimality analysis of new ad hoc tests that have yet to be developed.

The remainder of the paper proceeds as follows. In Section \ref{sec:intuition}, we provide the intuition underlying our numerical approach by illustrating how to find the weight function that justifies the two-sided $t$-test as a WAP-maximizing test and contrasting it with the point-wise power envelope that is not attained by the two-sided $t$-test.  Section \ref{sec: general framework} formalizes the general hypothesis testing framework we study, defines the numerical problem of finding weights that make an ad hoc test approximately WAP-maximizing and shows that the resulting WAP-maximizing test yields a “most favorable” approximate power envelope under a set of sufficient conditions.   Section \ref{section:numerical:implementation} describes our numerical procedure, detailing the inner loop algorithm that computes approximate WAP-maximizing tests and the outer loop algorithm that adjusts weights, along with practical considerations such as switching tests, thresholding, adding support points, and Monte Carlo smoothing.  We provide the theoretical convergence guarantees for these algorithms in Section \ref{sec:convergence theory},  establishing that our numerical method yields valid approximate WAP-maximizing tests and power envelopes.  In Section \ref{sec:applications}, we apply our results and numerical procedures to the two testing applications described above.  Mathematical proofs and implementation details for the applications are contained in the Appendix.

\section{Intuition and Motivation}\label{sec:intuition}

We begin by analyzing a simple canonical hypothesis testing example to impart intuition and motivate our approach to assessing the optimality of a hypothesis test.  Consider a two-sided test of the mean of a Gaussian random variable with known unit variance:~$H_0:\beta=0$ vs.\ $H_1:\beta\neq 0$ for $Y\sim\mathcal{N}(\beta,1)$.\footnote{The analysis of this section applies without loss of generality to all two-sided $t$-tests of the mean of a Gaussian random variable with known variance via a simple scale transformation.}  The standard level-$\alpha$ two-sided $t$-test that rejects when $|Y|$ exceeds $z_{1-\alpha/2}$ is well-known to be the uniformly most powerful test amongst all unbiased tests.  However, suppose that we would not like to confine our analysis to unbiased tests.  Since there is no uniformly most powerful test, we take the common approach of comparing the $t$-test to the point-wise power envelope for this testing problem.  Specifically, the point-wise power envelope is equal to the rejection probabilities of the collection of the most powerful point-wise tests of $H_0$ vs.\ $H_{\beta'}:\beta=\beta'$, as a function of a collection of $\beta'$ values.  By the Neyman-Pearson lemma, we know that each of these point-wise tests is the likelihood ratio test of $H_0$ vs.\ $H_{\beta'}$.

\begin{figure}[h]
  \begin{center}
    \includegraphics[width=100mm]{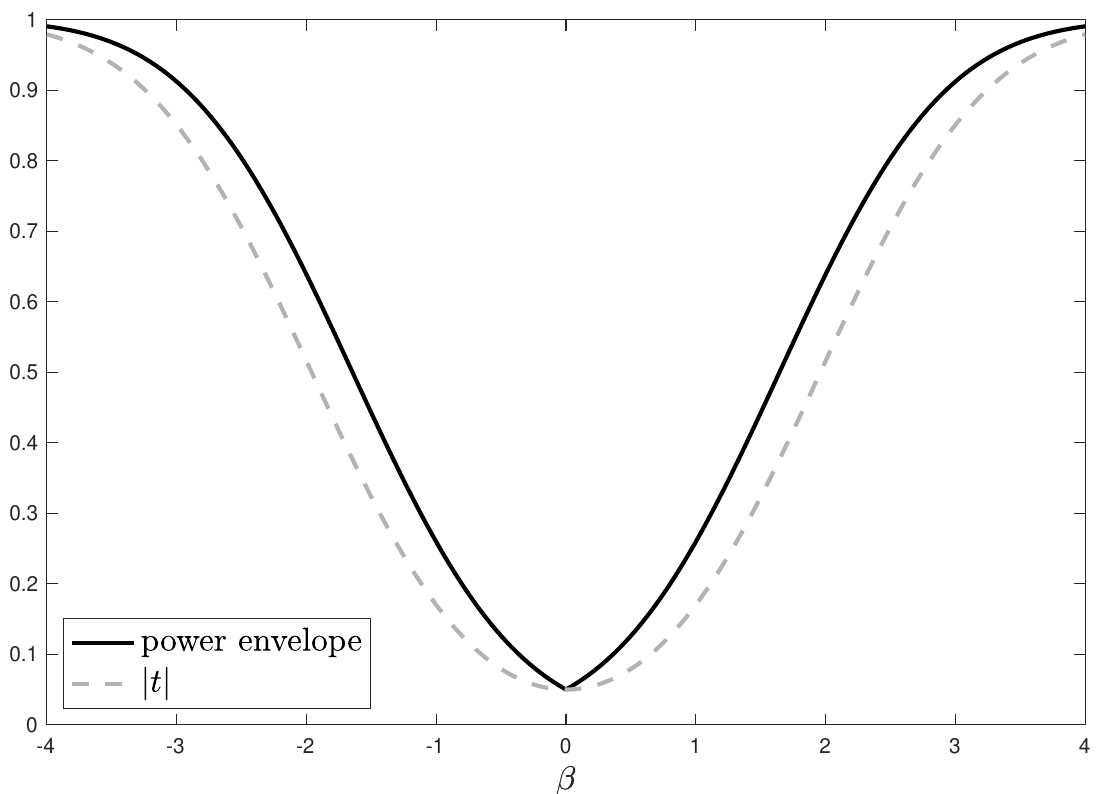}
     \caption{Power function of two-sided $t$-test and point-wise power envelope}\label{fig:2-sided Gaussian point-wise power bound}
    \end{center}
\end{figure}

Figure \ref{fig:2-sided Gaussian point-wise power bound} plots the power function of the two-sided $t$-test and the point-wise power envelope. Notably, the power function of the two-sided $t$-test lies substantially below the power envelope.  Using the point-wise power envelope could thus lead us to believe that the two-sided $t$-test is suboptimal.  However, note that the point-wise power envelope does not produce a power comparison that is necessarily attainable because it corresponds to the power of a \emph{collection} of point-optimal tests that are each optimal against a point alternative $H_{\beta'}$ with $\beta'\in\mathbb{R}\setminus\{0\}$, none of which correspond to the composite alternative hypothesis of interest $H_1$.  In other words, it is still possible for the two-sided $t$-test of the composite alternative $H_1$ to be optimal even though its power function lies below the power function of a collection of point-wise optimal tests of $H_{\beta'}$\textemdash we may simply be comparing its power function to an unattainable upper bound.

In the context of this testing problem, our proposed approach is to instead determine whether there exists a set of weights over the composite alternative space $\Theta_1=\mathbb{R}\setminus\{0\}$ such that the WAP of the two-sided $t$-test is well-approximated by the WAP of a WAP-maximizing test with this set of weights. If so, we may conclude that the two-sided $t$-test is ``nearly optimal'' for this set of weights.  To address this task, we discretize the composite alternative space $\Theta_1$ into support points $\bar\Theta_1\subset \Theta_1$ and introduce an algorithm that successively adjusts the weights on each support point to produce a WAP-maximizing test with WAP as close as possible to that of the two-sided $t$-test.  Surprisingly, we show below that the rejection probabilities of this WAP-maximizing test produce a power envelope for the two-sided $t$-test.  An important feature of this power envelope is that it corresponds to the rejection probabilities of a \emph{single} test and is therefore necessarily attainable, in contrast to the point-wise power envelope.  This power envelope is not only attainable but it is a ``most favorable'' power envelope in the sense that it is based upon a WAP-maximizing test with WAP as close as possible to that of the two-sided $t$-test itself.

For the sake of illustration, let us work with the coarse discretization $\bar\Theta_1=\{-1,1\}$.  The WAP-maximizing test of $H_0$ vs.\ the discretized composite alternative $\bar H_1:\beta\in \bar{\Theta}_1$ that weights $\beta=1$ by $\omega_1$ (and $\beta=-1$ by $1-\omega_1$) rejects when
\[
	LR(Y) = \frac{(1-\omega_1) f(Y;-1) + \omega_1 f(Y;1)}{f(Y;0)} > \text{cv}
\] 
for $\text{cv}$ satisfying $\mathbb{P}_{H_0}(LR(Y)>\text{cv}) = \alpha$, where $f(\cdot;\mu)$ denotes the density function of a $\mathcal{N}(\mu,1)$ random variable and $\mathbb{P}_{H_0}$ denotes the probability under $H_0$.  Our approach begins with some initial weight $\omega_1$, computes the power function of the WAP-maximizing test using this weight and adjusts the value of $\omega_1$ according to the difference between the power function of the $t$-test and the WAP-function at $\beta=1$:~if the power of the $t$-test lies below that of the WAP-maximizing test at $\beta=1$, we adjust $\omega_1$ downwards, putting more weight on $\beta=-1$, and vice versa.  We then repeat this procedure iteratively until the power function of the $t$-test lies weakly below that of the WAP-maximizing test at all points $\beta\in \bar\Theta_1$. If the power function of the $t$-test lies weakly below that of the WAP-maximizing test at all points $\beta\in \Theta_1$, the power function of this final WAP-maximizing test produces a power envelope that is based upon a \emph{single} test. If the power function of the $t$-test lies within a small value of the power function of this final WAP-maximizing test, we deem the $t$-test to be ``effectively optimal'' since its rejection probabilities nearly coincide with those of a test with known optimality.  The weights given in the final iteration tell us how the nearly optimal $t$-test weights points in the alternative space since they are the weights for which the corresponding WAP-maximizing test has WAP nearly equal to that of the $t$-test.  Otherwise, we deem the $t$-test to be suboptimal since its power function is dominated by that of another test. 

\begin{figure}[htbp]
    \centering
    
    \begin{subfigure}{0.45\textwidth}
        \includegraphics[width=\linewidth]{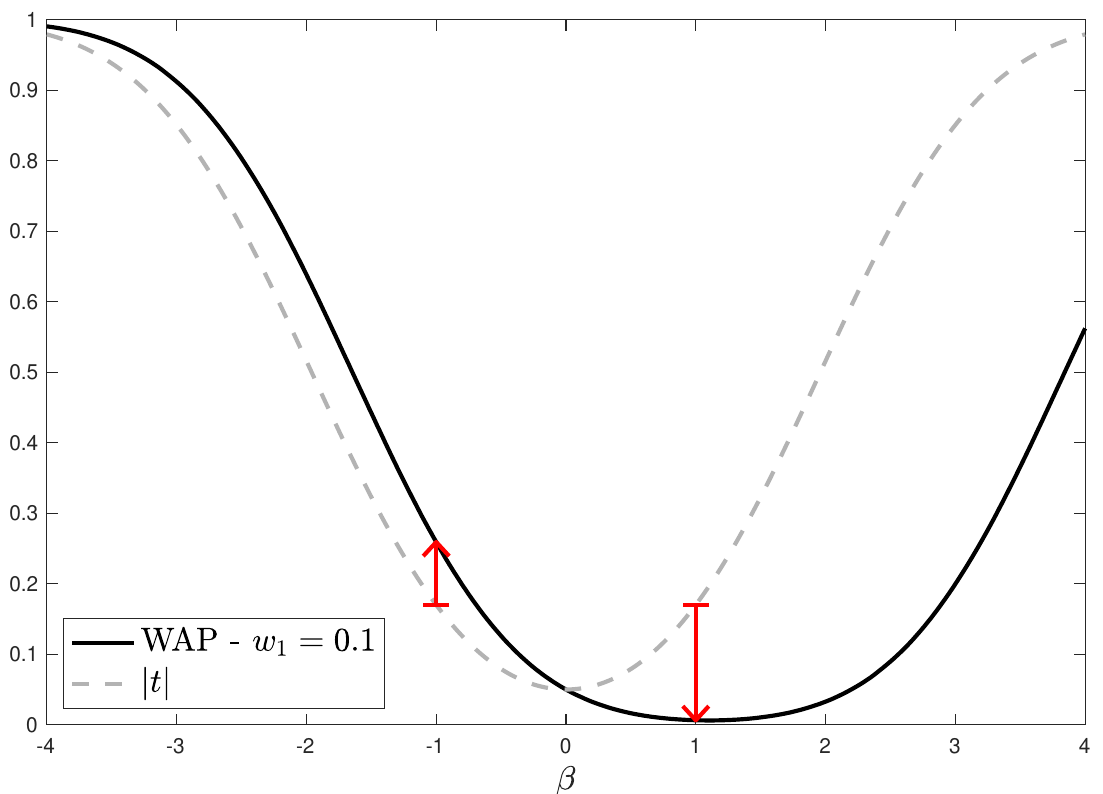}
        \caption*{$\omega_1=0.1$}
    \end{subfigure}
    \hfill
    \begin{subfigure}{0.45\textwidth}
        \includegraphics[width=\linewidth]{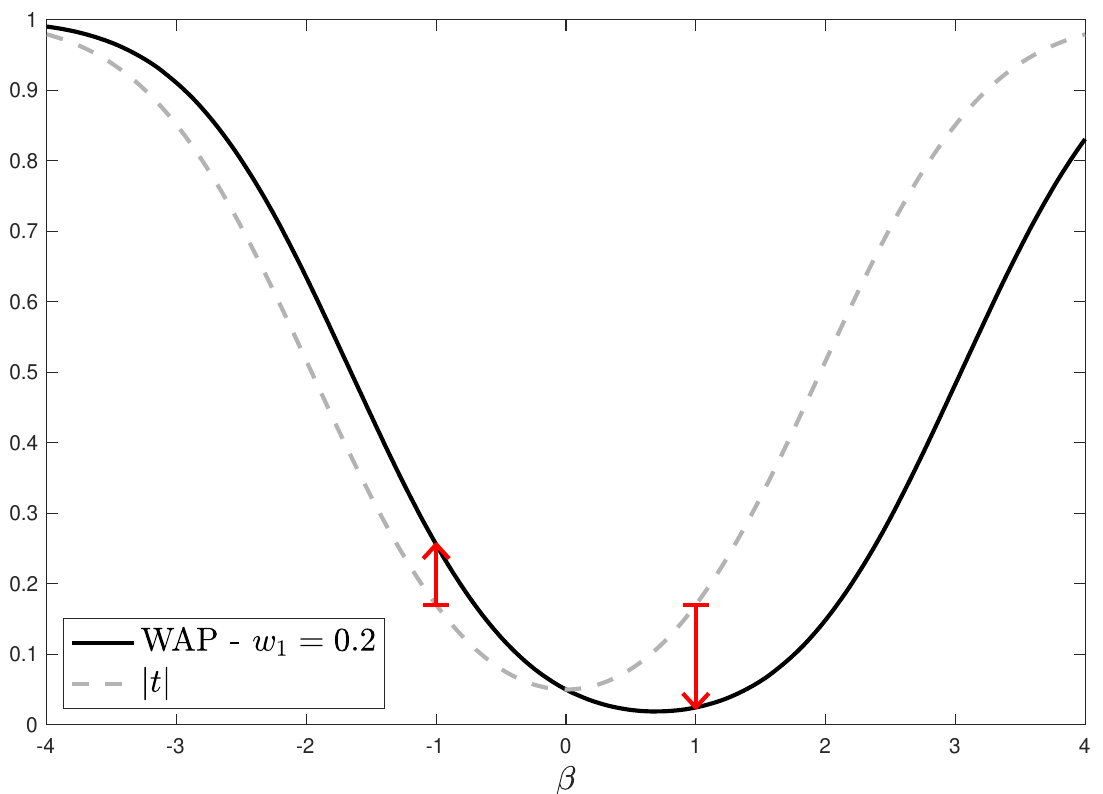}
        \caption*{$\omega_1=0.2$}
    \end{subfigure}
    
    \vspace{0.5cm} 
    
    \begin{subfigure}{0.45\textwidth}
        \includegraphics[width=\linewidth]{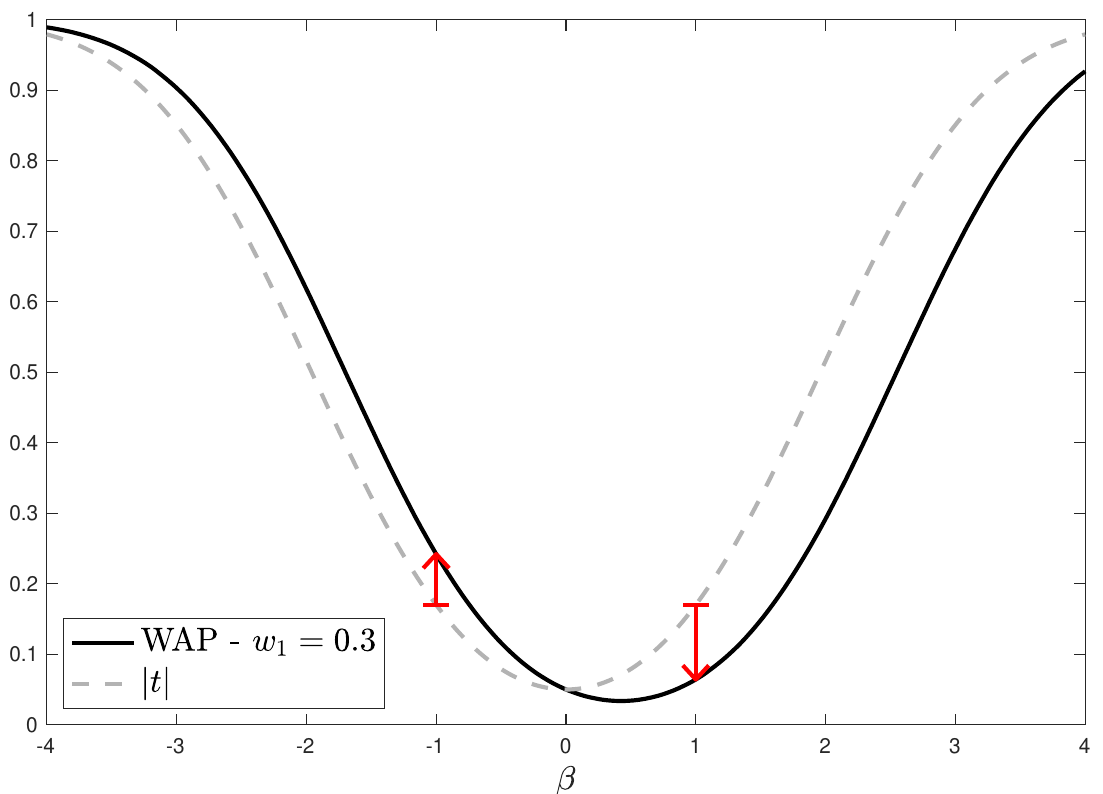}
        \caption*{$\omega_1=0.3$}
    \end{subfigure}
    \hfill
    \begin{subfigure}{0.45\textwidth}
        \includegraphics[width=\linewidth]{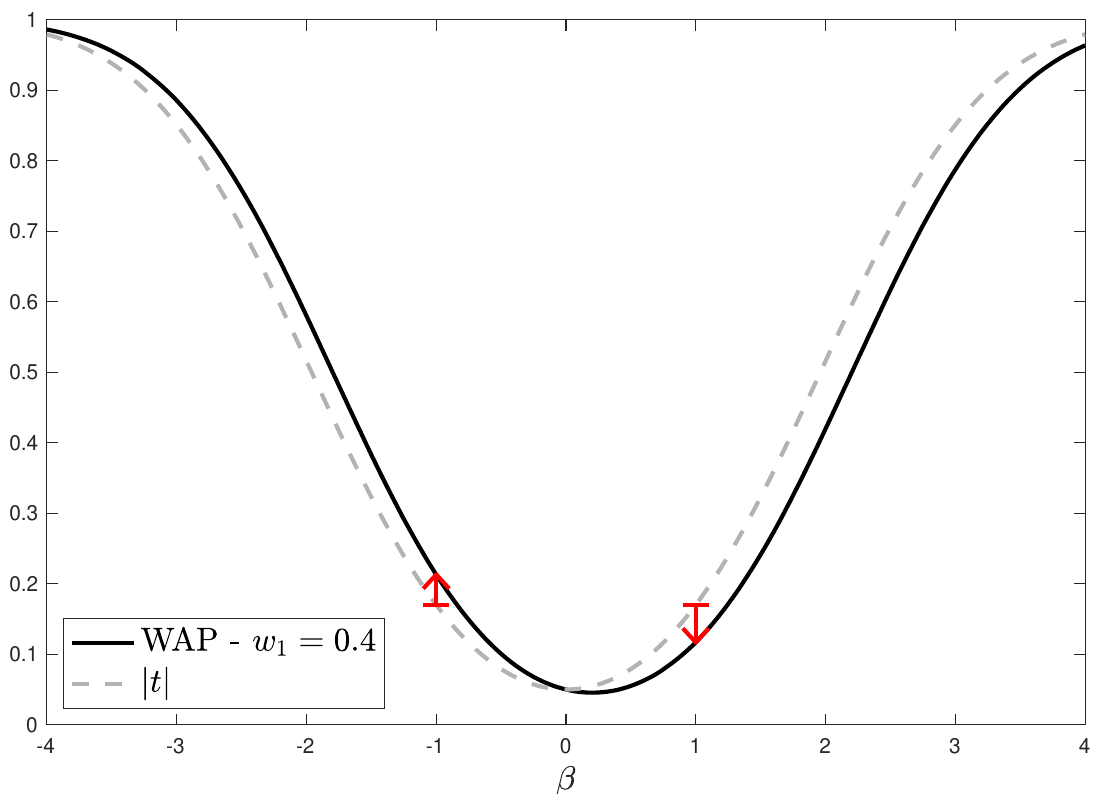}
        \caption*{$\omega_1=0.4$}
    \end{subfigure}
    
    \vspace{0.5cm}
    
    \begin{subfigure}{0.45\textwidth}
        \includegraphics[width=\linewidth]{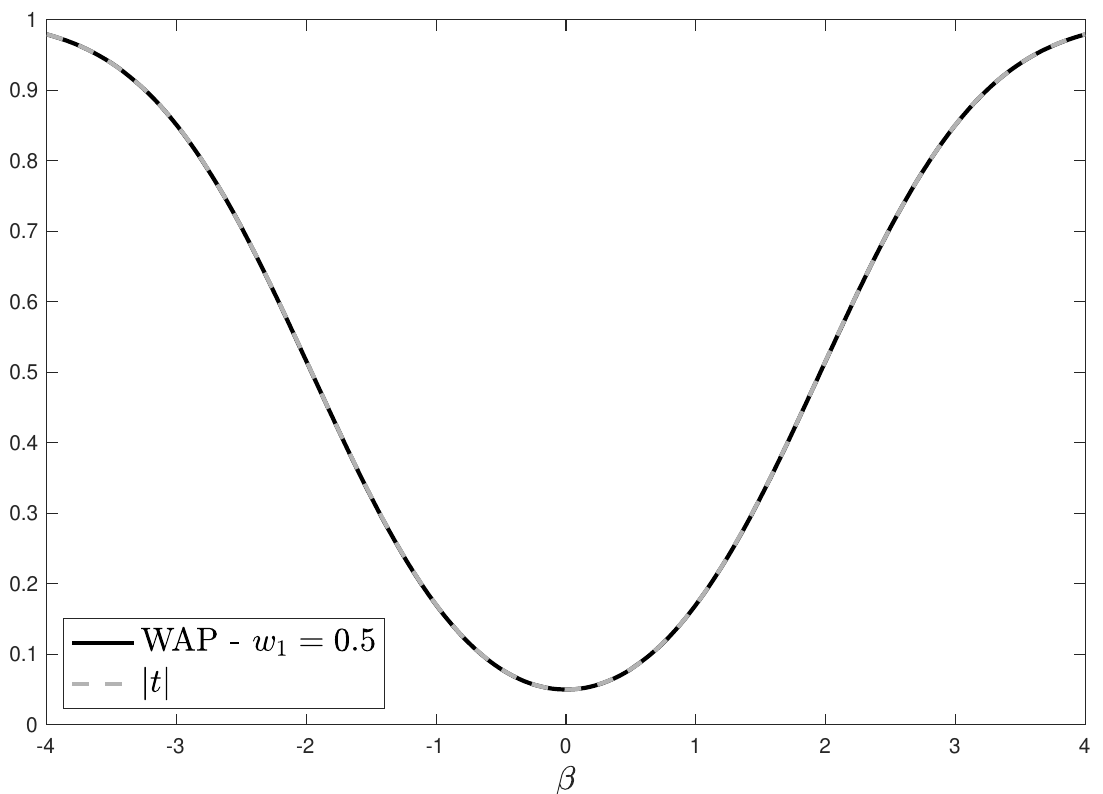}
        \caption*{$\omega_1=0.5$}
    \end{subfigure}
    
    \caption{Power functions for two-sided $t$-tests and WAP-maximizing tests with different weights}
    \label{fig: weight iterations for two-sided t-test}
\end{figure}

We provide the specifics of how the weight $\omega_1$ is adjusted at each iteration in the outer loop algorithm in Section \ref{sec:outer loop alg}, along with a theoretical justification for the algorithm in Section \ref{sec:convergence theory}, but the intuition is as follows.   Since any WAP-maximizing test is optimal for its set of weights by definition, we know (i) its power function cannot be dominated by that of the $t$-test on $\bar\Theta_1$ and (ii) if the $t$-test is itself WAP-optimal, there exists a set of weights for which its power function must match that of a WAP-maximizing test.  These two facts justify our iterative weighting adjustment:~at each iteration, we want to increase the weight at support points for which the WAP-maximizing test has lower power than the $t$-test and therefore decrease the weights at other support points.  Naturally, these decreased weights will be at points for which the WAP-maximizing test has greater power than the $t$-test.  The five panels of Figure \ref{fig: weight iterations for two-sided t-test} illustrate this re-weighting principle in five iterative steps, starting at $\omega_1=0.1$, for which $\omega_1$ is successively increased until the power function of the WAP-maximizing test coincides with that of the $t$-test.  The upward-pointing (downward-pointing) arrows indicate that the weight on the support point for the WAP-maximizing test needs to be adjusted downward (upward) to bring the power function of the WAP-maximizing test closer to lying (weakly) above that of the $t$-test. Since the power function of the WAP-maximizing test with weight $\omega_1=0.5$ in the final panel coincides with that of the two-sided $t$-test, we can say that the two-sided $t$-test is optimal in the class of all tests of $H_0$ vs.\ $H_1$ while no longer needing to constrain ourselves to the class of unbiased tests. 

\section{General Framework}\label{sec: general framework}

Having illustrated the intuition behind our approach in a simple problem, we now move to the general hypothesis testing framework of interest.

\subsection{Testing Problem}\label{sec: testing problem}

Suppose that we observe a random element $Y$ taking values in the metric space $\mathcal{Y}$ and that $Y$ has a probability density function $f_\theta(\cdot)$ relative to some sigma-finite measure $\nu$, where the parameter governing its distribution $\theta\in\Theta\subset\mathbb{R}^k$ is finite-dimensional.  We are interested in testing
\begin{equation}
	H_0: \theta \in \Theta_0 \text{ vs.\ } H_1: \theta \in \Theta_1, \label{eq: general hypotheses}
\end{equation}
where $\Theta_0, \Theta_1\subset\Theta$, $\Theta_0 \cap \Theta_1 = \emptyset$ and $\Theta_0$ is not a singleton.  This testing problem with a single observation $Y$ typically arises as the limiting problem under an asymptotic approximation to a finite-sample problem with many observations via a local asymptotic embedding corresponding to a limit experiment or by using the asymptotic equivalence approach of \cite{Muller:11}.  See \cite{EMW15} for a more detailed discussion.

A generic test of \eqref{eq: general hypotheses} is a measurable function $\varphi:\mathcal{Y}\mapsto [0,1]$ for which $\varphi(y)$ is the probability of rejecting $H_0$ upon observing the realization $Y = y$.  For the parameter value $\theta\in\Theta$, $\int\varphi f_\theta d\nu$ is thus equal to the rejection probability of the test when the true value of the parameter is $\theta$.  The starting point of our analysis is to suppose that we have an ad hoc test with correct size, a measurable function $\varphi_{ah}:\mathcal{Y}\mapsto [0,1]$ that is known to satisfy the (uniform) size constraint $\sup_{\theta\in\Theta_0}\int\varphi_{ah}f_\theta d\nu\leq \alpha$.  We would like to assess whether $\varphi_{ah}$ is ``nearly optimal'' among tests that control size.  For this problem to be nontrivial, we focus on tests of the hypothesis \eqref{eq: general hypotheses} for which no uniformly most powerful test is known to exist.

\subsection{Numerical Assessment of Near Optimality}\label{sec: WAP optimality}

As illustrated above in the simple context of a two-sided $t$-test for the mean of a Gaussian random variable with known variance, one could attempt to assess the optimality of $\varphi_{ah}$ by comparing it to the point-wise power envelope obtained by computing the rejection probabilities of a collection of level-$\alpha$ Neyman-Pearson tests indexed by $\theta'\in\Theta_1$ of $H_{0,\Lambda_{\theta'}}:$ the density of $Y$ is $\int f_\theta d\Lambda_{\theta'}(\theta)$ vs.\ $H_{\theta'}:\theta=\theta'$ given by
\[
	\varphi_{\Lambda_{\theta'},\theta'}(y) = \begin{cases}1 & \text{if } f_{\theta'}(y) > \text{cv} \int f_\theta(y) d\Lambda_{\theta'}(\theta) \\
							    \varkappa & \text{if } f_{\theta'}(y)  = \text{cv} \int f_\theta(y) d\Lambda_{\theta'}(\theta)  \\
							    0 & \text{if } f_{\theta'}(y)  < \text{cv} \int f_\theta(y) d\Lambda_{\theta'}(\theta)  \end{cases},
\]
for some cv $\geq 0$ and $0 \leq \varkappa \leq 1$ satisfying $\int\varphi_{\Lambda_{\theta'},\theta'}(\int f_\theta d\Lambda_{\theta'}(\theta)) d\nu= \alpha$, where $\Lambda_{\theta'}$ is the least-favorable probability distribution over $\Theta_0$ corresponding to the point alternative $H_{\theta'}$.\footnote{See \cite{LR05} for details on least-favorable distributions and \cite{AMS08}, \cite{EMW15} and \cite{AFBMOQST25} for details on computing approximations to them.}  Indeed, this approach has been applied in the literature, oftentimes after imposing additional side constraints such as similarity or invariance (e.g., \citealp{AMS06,AMS08}; \citealp{MM13,MM19}; \citealp{GKM19}; \citealp{AMY19}; \citealp{MO20}).  However, as in the simple example of Section \ref{sec:intuition}, comparison to this collection of Neyman-Pearson tests may produce an unattainable power envelope.

Instead, we propose to determine whether the WAP of $\varphi_{ah}$ is nearly equal to that of a \emph{single} WAP-maximizing test corresponding to some set of weights over $\Theta_1$ and is thus ``nearly optimal''.  More formally, we seek to determine whether there exists a weight function $\Omega$, a probability distribution with support on $\Theta_1$, such that $\varphi_{ah}$ nearly maximizes the WAP criterion (\citealp{Wal43})
\begin{equation}
	\text{WAP}(\varphi) = \int \left(\int \varphi f_\theta d\nu\right) d\Omega(\theta) \label{WAP criterion}
\end{equation}
within the set of level-$\alpha$ tests 
\[\Phi_\alpha\equiv\left\{\varphi:\mathcal{Y}\mapsto [0,1]:\varphi \text{ is measurable, }\sup_{\theta\in\Theta_0}\int \varphi f_\theta d\nu\leq \alpha\right\}.\]  
If a least-favorable probability distribution $\Lambda_\Omega$ over $\Theta_0$ corresponding to the simple alternative $H_{1,\Omega}:$ the density of $Y$ is $\int f_\theta d\Omega(\theta)$ exists, this WAP-maximizing test takes the familiar Neyman-Pearson form of a test of $H_{0,\Lambda_{\Omega}}:$ the density of $Y$ is $\int f_\theta d\Lambda_{\Omega}(\theta)$ vs.\ $H_{1,\Omega}$ given by
\begin{equation} \label{eq:NP test}
	\varphi_{\Lambda_{\Omega},\Omega}(y) = \begin{cases}1 & \text{if } \int f_\theta(y) d\Omega(\theta) > \text{cv}_{\Omega} \int f_\theta(y) d\Lambda_{\Omega}(\theta) \\
							    \varkappa_\Omega & \text{if } \int f_\theta(y) d\Omega(\theta)  = \text{cv}_{\Omega} \int f_\theta(y) d\Lambda_{\Omega}(\theta)  \\
							    0 & \text{if } \int f_\theta(y) d\Omega(\theta)  < \text{cv}_{\Omega} \int f_\theta(y) d\Lambda_{\Omega}(\theta). \end{cases},
\end{equation}
Here, $\text{cv}_{\Omega}\geq 0$ and $0 \leq \varkappa_\Omega \leq 1$ are defined to satisfy $\int\varphi_{\Lambda_{\Omega},\Omega}(\int f_\theta d\Lambda_{\Omega}(\theta)) d\nu= \alpha$.  As noted by \cite{EMW15}, even if this least-favorable distribution does not exist, we can find an ``approximate least-favorable distribution'' to approximate the test the maximizes \eqref{WAP criterion}.

It is not typically possible to determine the existence of a weight function $\Omega$ for which an ad hoc test $\varphi_{ah}$ maximizes the WAP criterion WAP$(\varphi)$ analytically.  We thus propose a numerical approach that computes the weights that make the WAP of a WAP-maximizing test as close as possible to the WAP of the ad hoc test \emph{for those weights} over a discretization of the alternative space.  Specifically, let $\bar\Theta_1=(\theta_1,\ldots,\theta_{M_1})\subset\Theta_1$ denote a finite set of support points for the alternative space with $M_1$ elements and $\bar\Omega=(\omega_1,\ldots,\omega_{M_1})\in \Delta_{M_1}$ denote a probability distribution over those support points, where $\Delta_{M_1}$ is the $M_1$-dimensional unit simplex.  Formally, the numerical problem we wish to solve is given by
\begin{align}
    \inf_{\bar\Omega\in \Delta_{M_1}} \sup_{\varphi\in \Phi_\alpha} \sum_{j=1}^{M_1} \omega_j \int(\varphi - \varphi_{ah}) f_{\theta_j} d\nu. \label{eq:discrete power optimization}
\end{align}
A test $\varphi^*\in\Phi_\alpha$ that solves \eqref{eq:discrete power optimization} is a WAP-maximizing test with corresponding WAP as close as possible to that of $\varphi_{ah}$ since 
\begin{equation}
    \sup_{\varphi\in \Phi_\alpha} \sum_{j=1}^{M_1} \omega_j \int(\varphi - \varphi_{ah}) f_{\theta_j} d\nu=\sup_{\varphi\in \Phi_\alpha} \sum_{j=1}^{M_1} \omega_j \int\varphi f_{\theta_j} d\nu  - \sum_{j=1}^{M_1} \omega_j \int \varphi_{ah} f_{\theta_j} d\nu. \label{eq:discrete WAP-max}
\end{equation}
A natural byproduct of our approach is that a weight function $\bar\Omega$ solving \eqref{eq:discrete power optimization} provides valuable information about the power properties of the ad hoc test:~higher weight over a set of points in $\bar\Theta_1$ indicates that the test prioritizes rejection in that set relative to other points in $\bar\Theta_1$.  

\subsection{Using WAP-Maximizing Tests to Produce an Approximate Power Envelope}

In addition to determining whether $\varphi_{ah}$ nearly maximizes a WAP criterion for some set of weights via \eqref{eq:discrete power optimization}, our numerical approach has the added benefit of being able to produce an approximate power envelope (APE) for $\varphi_{ah}$, whether or not it is WAP-maximizing. This power envelope is again obtained from a \emph{single} test and is therefore necessarily attainable, in contrast to power envelopes derived from a collection of tests.  To see this, first note that the rejection probabilities over $\Theta_1$ of the test that solves
\[\sup_{\varphi\in\Phi_\alpha}\inf_{\theta\in\Theta_1}\int(\varphi-\varphi_{ah})f_\theta d\nu\]
constitute a power envelope for $\varphi_{ah}\in\Phi_\alpha$ since
\[\sup_{\varphi\in\Phi_\alpha}\inf_{\theta\in\Theta_1}\int(\varphi-\varphi_{ah})f_\theta d\nu\geq\inf_{\theta\in\Theta_1}\int(\varphi_{ah}-\varphi_{ah})f_\theta d\nu=0. \]
For an appropriately chosen $\bar\Theta_1=(\theta_1,\ldots,\theta_{M_1})\subset\Theta_1$\textemdash see Section \ref{sec:add:supp:pts} for details, we can then view the rejection probabilities of the test that solves
\begin{equation}
\sup_{\varphi\in\Phi_\alpha}\min_{j=1,\ldots,M_1}\int(\varphi-\varphi_{ah})f_{\theta_j} d\nu \label{eq:maxmin problem}
\end{equation}
as an APE for $\varphi_{ah}$ since it constitutes a power envelope for $\varphi_{ah}$ over $\bar \Theta_1$ by definition.  In Theorem \ref{thm:WAP-max power bound equivalence} below, we show that the value of the maximin problem in \eqref{eq:maxmin problem} is equal to the value of the minimax problem in \eqref{eq:discrete power optimization} and any test that solves \eqref{eq:maxmin problem} also solves \eqref{eq:discrete power optimization}.  Since we show how to approximate the solution to \eqref{eq:discrete power optimization} in Sections \ref{section:numerical:implementation} and \ref{sec:convergence theory} below, this latter result provides a practical strategy for obtaining an APE for $\varphi_{ah}$:~solve \eqref{eq:discrete power optimization} and numerically check whether the solution yields an APE for $\varphi_{ah}$.  Absent further structure on the problem, this strategy is not guaranteed to produce an APE but it can still provide a useful guide for obtaining an APE in practice.  In Theorem \ref{thm:N-P test power envelope} below, we impose additional structure on the problem that is sufficient for guaranteeing that this strategy indeed produces an APE.   


Let
\[
    \Phi^\ast = \argmax_{\varphi\in \Phi_\alpha} \min_{j=1,\ldots,M_1}\int(\varphi-\varphi_{ah})f_{\theta_j} d\nu 
\]
and
\[
    \Delta_{M_1}^\ast = \argmin_{\bar\Omega\in \Delta_{M_1}} \sup_{\varphi\in \Phi_\alpha} \sum_{j=1}^{M_1} \omega_j \int(\varphi - \varphi_{ah}) f_{\theta_j} d\nu.
\]
We can now state our first result.

\begin{theorem}\label{thm:WAP-max power bound equivalence}
For $\bar\Omega=(\omega_1,\ldots,\omega_{M_1})\in \Delta_{M_1}$ and any $\alpha\in (0,1)$, 
\[\sup_{\varphi\in\Phi_\alpha}\min_{j=1,\ldots,M_1}\int(\varphi-\varphi_{ah})f_{\theta_j} d\nu =\inf_{\bar\Omega\in \Delta_{M_1}} \sup_{\varphi\in \Phi_\alpha} \sum_{j=1}^{M_1} \omega_j \int(\varphi - \varphi_{ah}) f_{\theta_j} d\nu,\]
$\Phi^\ast \neq \emptyset$ and $\Delta_{M_1}^\ast \neq \emptyset$. Furthermore, for any $\varphi^* \in \Phi^*$ and any $\bar \Omega^*=(\omega_1^*,\ldots,\omega_{M_1}^*) \in \Delta_{M_1}^\ast$, we have
\begin{equation} \label{eq:WAP:maximizer:is:maximin}
    \sum_{j=1}^{M_1} \omega_j^* \int(\varphi^* - \varphi_{ah}) f_{\theta_j} d\nu
    = \sup_{\varphi\in \Phi_\alpha} \sum_{j=1}^{M_1} \omega_j^* \int(\varphi - \varphi_{ah}) f_{\theta_j} d\nu.
\end{equation}
\end{theorem}

In words, equation \eqref{eq:WAP:maximizer:is:maximin} states that any $\varphi^* \in \Phi^*$, i.e., any test that solves the maximin problem in \eqref{eq:maxmin problem}, is WAP-maximizing with respect to any $\bar \Omega^* \in \Delta_{M_1}^*$. Therefore, if one follows the strategy mentioned above and finds that a test that solves \eqref{eq:discrete power optimization} yields an APE for $\varphi_{ah}$, this strategy comes with the added benefit that the APE is ``most-favorable'' for $\varphi_{ah}$ in the sense that it corresponds to the rejection probabilities of a WAP-maximizing test with WAP as close as possible to that of $\varphi_{ah}$ over $\bar\Theta_1$.


It can be shown that Theorem \ref{thm:WAP-max power bound equivalence} further implies that if the least-favorable distribution $\Lambda_{\bar \Omega^*}$ exists for all $\bar \Omega^* \in \Delta_{M_1}^\ast$, then any test $\varphi^* \in \Phi^*$\textemdash which serves as a power envelope for $\varphi_{ah}$ on $\bar\Theta_1$\textemdash is equal to a Neyman-Pearson test $\varphi_{\Lambda_{\bar \Omega^*},\bar \Omega^*}$ for some $\bar \Omega^* \in \Delta_{M_1}^\ast$, except possibly on the event:\footnote{This follows from Lemma \ref{pot_power_fun_inner_loop_new} under the assumption that $\Lambda^\ast(\Theta_0) > 0$ for all $\Lambda^\ast \in \mathcal{M}_0^\ast$.}
\begin{align}
    \biggl\{y\in \mathcal{Y} : \sum_{j=1}^{M_1} \omega_j^* f_{\theta_j}  = \mathrm{cv}_{\bar \Omega^*} \int f_\theta(y) d\Lambda_{\bar \Omega^*} (\theta)\biggr\}.\label{ass_uniqueness}
\end{align}
We impose the commonly-satisfied sufficient condition that this event has $\nu$-measure zero for all $\bar \Omega^*$ in Theorem \ref{thm:N-P test power envelope} below and formally show that any $\varphi^* \in \Phi^*$ is equal to a Neyman-Pearson test $\varphi_{\Lambda_{\bar \Omega^*},\bar \Omega^*}$ $\nu\text{-almost everywhere}$ (a.e.), and this Neyman-Pearson test is unique $\nu\text{-a.e.}$  This uniqueness result trivially implies that the test that solves \eqref{eq:discrete power optimization} also solves \eqref{eq:maxmin problem} so that the test obtained by solving the minimax problem according to our algorithms in Section \ref{section:numerical:implementation} readily provides a most-favorable APE for $\varphi_{ah}$.

\begin{theorem}\label{thm:N-P test power envelope}
Let $\alpha\in (0,1)$.  For all $\bar \Omega^\ast\in \Delta_{M_1}^\ast$, suppose that the least favorable distribution $\Lambda_{\bar \Omega^*}$ over $\Theta_0$ exists and that
\eqref{ass_uniqueness} has $\nu$-measure zero. Then, for any $\bar \Omega^\ast\in \Delta_{M_1}^\ast$, $\varphi_{\Lambda_{\bar \Omega^\ast},\bar \Omega^\ast}$ is the $\nu$-a.e.\ unique maximizer of 
\[
    \sup_{\varphi\in \Phi_\alpha} \sum_{j=1}^{M_1} \omega_j^* \int(\varphi - \varphi_{ah}) f_{\theta_j} d\nu.
\]
Furthermore, for any $\varphi^\ast_1,\varphi^\ast_2 \in \Phi^*$ and any $\bar \Omega^\ast_1, \bar \Omega^\ast_2 \in \Delta_{M_1}^\ast$, we have
\[
\varphi^\ast_1 = \varphi^\ast_2 = \varphi_{\Lambda_{\bar \Omega^\ast_1},\bar \Omega^\ast_1}= \varphi_{\Lambda_{\bar \Omega^\ast_2},\bar \Omega^\ast_2} \ \nu\text{-a.e.}
\]
\end{theorem}

The condition that \eqref{ass_uniqueness} has $\nu$-measure zero typically holds when $Y$ is an absolutely continuous random vector since \eqref{ass_uniqueness} is typically a lower-dimensional submanifold of $\mathcal Y$.  The existence of a least favorable distribution has been established under weak conditions for a Euclidean sample space $\mathcal Y$ when $f_\theta$ is continuous in $\theta$ and $\Theta_0$ is a closed Borel set in a finite-dimensional Euclidean space.  See \cite{LR05} and references therein.  

\section{Numerical Implementation} \label{section:numerical:implementation}

To numerically approximate the weight function that justifies an ad hoc test $\varphi_{ah}$ in terms of WAP or produce a power function that dominates it, we aim to solve \eqref{eq:discrete power optimization}.  Our numerical algorithm for solving \eqref{eq:discrete power optimization} is composed of an inner loop and an outer loop.  The inner loop computes an approximation to a WAP-maximizing test $\phi_{\Lambda_\Omega,\Omega}$ \emph{for a given weight function} $\Omega$ in the spirit of \cite{MM13}, \cite{EMW15} and \cite{AFBMOQST25}.\footnote{\cite{GH24} also briefly note that their numerical algorithm for approximating minimax regret treatment rules could also potentially be modified to numerically approximate a WAP-maximizing test for a given weight function.}  The outer loop approximates a weight function $\bar\Omega$ that solves \eqref{eq:discrete power optimization} using the inner loop as input at each step since it involves searching over WAP-maximizing tests via the relation \eqref{eq:discrete WAP-max}.  In Section \ref{sec:convergence theory}, we provide theoretical convergence results justifying the use of our algorithm for solving \eqref{eq:discrete power optimization}.

\subsection{Inner Loop for Computing Approximate WAP-Maximizing Test}\label{sec:inner loop alg}

Discretize the null parameter space $\Theta_0$ into a finite set of support points $\bar\Theta_0=\{\tilde\theta_1,\ldots,\tilde\theta_{M_0}\}\subset\Theta_0$ with $M_0$ elements.  In light of \eqref{WAP criterion} (and Fubini's Theorem), we aim to solve the following discretized optimization problem:
\begin{equation} 
    \max_{\varphi\in \Phi} \int \varphi g d\nu \quad \text{ s.t. } \int \varphi f_{\tilde\theta_i} d\nu \le \alpha, \quad \text{for } i = 1, \ldots, M_0, \label{eq:discrete null problem}
\end{equation}
where $\Phi = \{ \varphi: \mathcal{Y} \to [0,1]: \varphi \text{ is measurable}\}$ and $g(y)=\int f_\theta(y)d\Omega(\theta)$ for a given weight function $\Omega$.  Its dual problem is given by
\begin{align*}
    \min_{\widetilde\Lambda \ge 0}\text{ }\max_{\varphi \in \Phi} \int \varphi gd\nu - \sum_{i=1}^{M_0} \tilde\lambda_i \biggl( \int \varphi f_{\tilde\theta_i} d\nu - \alpha\biggr),
\end{align*}
where $\widetilde\Lambda=(\tilde\lambda_1,\ldots,\tilde\lambda_{M_0})$.  It is not hard to see that Slater's condition is satisfied in this setting and therefore strong duality holds between these two problems.\footnote{See Theorem 1 in Chapter 8.3 in \cite{luenberger1997optimization}.} This implies that for any solution $\widetilde\Lambda^\ast$ of the dual problem, any solution $\tilde\varphi_{\widetilde\Lambda^\ast}$ of
\begin{align*}
    \tilde\phi(\widetilde\Lambda^\ast) := \max_{\varphi \in \Phi} \int \varphi gd\nu - \sum_{i=1}^{M_0} \tilde\lambda_i^\ast \biggl( \int \varphi f_{\tilde\theta_i} d\nu - \alpha\biggr)
\end{align*}
is also a solution of the primal problem \eqref{eq:discrete null problem}. We can operationalize this observation via the following simple result.

\begin{proposition}\label{NP_thm_KKT_form}
    For any $\widetilde\Lambda \ge 0$,
    \begin{align*}
        \tilde\varphi_{\widetilde\Lambda} \in \arg\max_{\varphi \in \Phi} \int \varphi g d\nu 
        - \sum_{i=1}^{M_0} \tilde\lambda_i \biggl( \int \varphi f_{\tilde\theta_i} d\nu - \alpha\biggr)
    \end{align*}
    for 
    \begin{align*}
        \tilde\varphi_{\widetilde\Lambda} (y) = \begin{cases}
            1 & \text{, if } g(y) \ge \sum_{i=1}^{M_0} \tilde\lambda_i f_{\tilde\theta_i}(y)\\
            0 & \text{, if } g(y) < \sum_{i=1}^{M_0} \tilde\lambda_i f_{\tilde\theta_i}(y).
        \end{cases} 
    \end{align*}
\end{proposition}

Since $\tilde\varphi_{\widetilde\Lambda^\ast}$ is given in closed form as soon as we know $\widetilde\Lambda^\ast$, it is sufficient to solve the dual problem in order to find a tractable solution to the primal problem \eqref{eq:discrete null problem}. Also, note that given the above, we have
\begin{equation}
    \tilde\phi(\widetilde\Lambda) = \int \tilde\varphi_{\widetilde\Lambda} g d\nu 
    - \sum_{i=1}^{M_0} \tilde\lambda_i \biggl( \int \tilde\varphi_{\widetilde\Lambda} f_{\tilde\theta_i} d\nu - \alpha\biggr). \label{eq:reformulated dual problem}
\end{equation}
This observation motivates the following algorithm for approximating $\tilde\varphi_{\widetilde\Lambda^\ast}$.

\begin{enumerate}
\item \underline{Initialization:}
\begin{enumerate}
\item Choose $\widetilde\Lambda^{(0)} = (\tilde\lambda_1^{(0)}, \dots, \tilde\lambda_{M_0}^{(0)})\geq 0$ and $\{\tilde h_k\}_{k=0}^\infty$ satisfying
        \begin{align*}
            \tilde h_k > 0 \quad \text{for } k=0,1,\ldots, \quad \tilde h_k \to 0 \quad \text{as } k\to\infty
            \quad \text{and} \quad \sum_{k=0}^\infty \tilde h_k = \infty.
        \end{align*}
\item Compute $\tilde\varphi_{\widetilde\Lambda^{(0)}}$ and $\tilde\phi(\widetilde\Lambda^{(0)})$.
\end{enumerate}
\item \underline{$k$th iteration:}
\begin{enumerate}
\item Compute $\tau = (\tau_1,\dots, \tau_{M_0})$, where
        \begin{align*}
            \tau_i = \int \tilde\varphi_{\widetilde\Lambda^{(k)}} f_{\tilde\theta_i} d\nu - \alpha.
        \end{align*}
\item Update $\widetilde\Lambda^{(k+1)} = (\tilde\lambda_1^{(k+1)}, \dots, \tilde\lambda_{M_0}^{(k+1)})$:
        \begin{align*}
            \tilde\lambda_i^{(k+1)} = \biggl( \tilde\lambda_i^{(k)} + \tilde h_k \frac{\tau_i}{\lVert \tau_i\rVert _2} \biggr) _+ \quad \text{for } i=1,\ldots,M_0, 
        \end{align*}
        where $(a)_+ = \max\{a,0\}$ for $a \in \mathbb{R}$.
\item Compute $\tilde\varphi_{\widetilde\Lambda^{(k+1)}}$ and $\tilde\phi(\widetilde\Lambda^{(k+1)})$.
\end{enumerate}
\end{enumerate}

The intuition for this algorithm is essentially the same as that of \cite{EMW15}.  However, in contrast to the algorithm in \cite{EMW15}, our algorithm carries proven convergence guarantees\textemdash see Section \ref{sec:convergence theory}.\footnote{\cite{AFBMOQST25} show that a slight modification to the algorithm of \cite{EMW15} produces an algorithm with proven convergence guarantees.}

For $k$ large enough, a $\tilde\varphi_{\widetilde\Lambda^{(k)}}$ approximately solves the original optimization problem \eqref{eq:discrete null problem} in a sense we make precise in Section \ref{sec:convergence theory}.  By Lemma 1 of \cite{EMW15}, if $\sup_{\theta\in\Theta_0}\int\tilde\varphi_{\widetilde\Lambda^{(k)}}f_\theta d\nu\leq \alpha$, then $\tilde\varphi_{\widetilde\Lambda^{(k)}}$ attains an upper bound on the WAP for the weight function $\Omega$ for any test $\varphi\in\Phi_\alpha$, no matter how large $k$ is.  This size constraint needs to be checked numerically but approximately holds by design if $\bar\Theta_0\subset\Theta_0$ is rich enough.  Finally, to map any test $\tilde\varphi_{\widetilde\Lambda}$ given in Lagrange multiplier form back to the Neyman-Pearson form \eqref{eq:NP test}, simply set  $\text{cv}_{\Omega}=\sum_{i=1}^{M_0}\tilde\lambda_i$, $\varkappa_\Omega=1$ and $\Lambda=(\lambda_1,\ldots,\lambda_{M_0})\in\Delta_{M_0}$ with $\lambda_i=\tilde\lambda_i/\text{cv}_{\Omega}$ for $i=1,\ldots,M_0$.

\subsection{Outer Loop Algorithm for Computing Weight Function}\label{sec:outer loop alg}

Let
\begin{align*}
    \phi(\bar\Omega) = \sup_{\varphi\in \Phi_\alpha} \sum_{j=1}^{M_1} \omega_j \int(\varphi - \varphi_{ah}) f_{\theta_j} d\nu
    \end{align*}
    and
    \begin{align*}
    \varphi_{\bar\Omega}^*\in\arg\max_{\varphi\in \Phi_\alpha} \sum_{j=1}^{M_1} \omega_j \int(\varphi - \varphi_{ah}) f_{\theta_j} d\nu.
\end{align*}
The function $\phi$ may fail to be fully differentiable but is convex, therefore enabling us to work with the following projected subgradient descent algorithm to solve \eqref{eq:discrete power optimization}.

\begin{enumerate}
\item \underline{Initialization:}
\begin{enumerate}
\item Choose $\bar\Omega^{(0)} = (\omega_1^{(0)}, \dots, \omega_{M_1}^{(0)})\in\Delta_{M_1}$ and $\{h_k\}_{k=0}^\infty$ satisfying
\begin{align*}
             h_k > 0 \quad \text{for } k=0,1,\ldots, \quad  h_k \to 0 \quad \text{as } k\to\infty
            \quad \text{and} \quad \sum_{k=0}^\infty  h_k = \infty.
        \end{align*}
\item Compute $\phi(\bar\Omega^{(0)})$ and 
        \begin{align*}
            \varphi_{\bar\Omega^{(0)}}^*\in \arg\max_{\varphi \in \Phi_\alpha} \sum_{j=1}^{M_1} \omega_j^{(0)} \int(\varphi - \varphi_{ah}) f_{\theta_j} d\nu
        \end{align*}
    using the inner loop algorithm.
\end{enumerate}
\item \underline{$k$th iteration:}
\begin{enumerate}
\item Compute $\gamma = (\gamma_1,\dots, \gamma_{M_1})$, where
        \begin{align*}
            \gamma_i = \int(\varphi_{\bar\Omega^{(k)}}^* - \varphi_{ah}) f_{\theta_i} d\nu, \quad \text{for } i = 1, \dots, M_1.
        \end{align*}
\item Update $\bar\Omega^{(k+1)} = (\omega_1^{(k+1)}, \dots, \omega_{M_1}^{(k+1)})$:
        \begin{align*}
            \omega_i^{(k+1)} = \pi_\Delta\biggl( \omega_i^{(k)} - h_k \frac{\gamma_i}{\lVert \gamma_i\rVert _2} \biggr) \quad \text{for } i=1,\ldots,M_1, 
        \end{align*}
        where $\pi_\Delta$ denotes the Euclidean projection onto the unit simplex $\Delta_{M_1}$.
\item Compute $\phi(\bar\Omega^{(k+1)})$ and 
        \begin{align*}
            \varphi_{\bar\Omega^{(k+1)}}^* \in \arg\max_{\varphi \in \Phi_\alpha} \sum_{j=1}^{M_1} \omega_j^{(k+1)} \int(\varphi - \varphi_{ah}) f_{\theta_j} d\nu
        \end{align*}
    using the inner loop algorithm.
        
\end{enumerate}
\end{enumerate}

The intuition underlying the algorithm is simple.  Since our goal is to obtain a power envelope for $\varphi_{ah}$ from a WAP-maximizing test, we aim to find weights $\bar\Omega$ such that the power of $\varphi_{\bar\Omega}^*$ is (weakly) greater than that of $\varphi_{ah}$ at all support points in $\bar\Theta_1$.  So at each iteration $k$, we aim to increase the weight given to support points for which $\varphi_{\bar\Omega^{(k)}}^*$ currently has lower power than $\varphi_{ah}$.  Since the weights must add to one, we therefore must decrease the weight given to other support points.  Naturally, we do so at points for which $\varphi_{\bar\Omega^{(k)}}^*$ currently has higher power than $\varphi_{ah}$.  Step 2(a) computes the support points for which we aim to increase (decrease) the corresponding weights while Step 2(b) computes the weight adjustment.

For $k$ large enough, a $\bar\Omega^{(k)}$ approximately solves the original optimization problem \eqref{eq:discrete power optimization} in a sense we make precise in Section \ref{sec:convergence theory}.  Step 2(b) of the algorithm relies on a Euclidean projection onto the unit simplex $\Delta_{M_1}$.  This is a quadratic programming problem and there are efficient computational algorithms in the literature designed for solving this problem (e.g.,~\citealp{wang2013simplexefficient}).


\subsection{Implementation Details} \label{sec:implementation}

In this section, we provide precise recommendations on how to implement our numerical procedure.  We follow these recommendations ourselves when analyzing the applications in Section \ref{sec:applications}.

\subsubsection{Switching Tests} \label{sec:switching:tests}
We follow \cite{EMW15} and implement switching tests when the parameter space for the nuisance parameter is unbounded, albeit with an additional related objective.  Tests that do not (approximately) reduce to a standard test in the ``standard'' part of the parameter space, typically characterized by large values of nuisance parameters, tend to significantly sacrifice power in the ``standard'' part of the parameter space, a conclusively undesirable property (see Section 4 of \citealp{EMW15}).  Therefore, we do not wish to limit the scope of our analysis on test optimality to WAP-maximizing tests that are not able to ``switch'' to standard tests in the ``standard'' parts of the parameter space.  Indeed, ad hoc tests are typically purposefully designed to reduce to standard tests that are known to be optimal in some sense in the ``standard'' part of the parameter space for this very reason.  For example, in the linear instrumental variables (IV) model, the CLR test reduces to the two-sided $t$-test when the concentration parameter is large.  Since our optimality-assessment procedure relies upon approximating WAP-maximizing tests on a finite set of support points $\bar\Theta_1\subset\Theta_1$, comparing to WAP-maximzing tests that do no allow for ``switching'' would inherently disadvantage an ad hoc test that reduces to a standard test in the ``standard'' part of the parameter space because the WAP-maximizing test would place zero weight over any region in $\Theta_1$ outside of $\bar\Theta_1$.  Due to the typical noncompactness of the parameter spaces $\Theta_0$ and $\Theta_1$, there are also numerical benefits to focusing on switching tests, for which we also refer the interested reader to Section 4 of \cite{EMW15}.


Since an ad hoc test can immediately be seen as suboptimal when it does not reduce to a known ``standard best test'' in the ``standard'' part of the parameter space, we focus here on ad hoc tests that do.  Section 4.1 of \cite{EMW15} formalizes how the ``standard'' part of the parameter space can be characterized in terms of large values of a parameter $\delta$.  Let $\delta_\text{S}$ be the point where the ``standard'' part of the parameter space begins in the sense that for $\delta > \delta_\text{S}$ the ad hoc test and the standard test have essentially the same rejection profile.  Let $D(Y)$ and $\delta_\text{SP}$ denote a statistic and a ``switching point'' such that with probability very close to zero, $D(Y)>\delta_\text{SP}$ whenever $\delta\leq \delta_\text{S}$.  The motivation for this choice is that we only want the test to which we compare $\varphi_{ah}$ to switch to a standard test $\varphi_S$ that has the best rejection profile in the standard part of the parameter space when we know that the ad hoc test has essentially the same rejection profile, enabling us to analyze its optimality properties outside of this region.  In practice, all this amounts to changing $\tilde\varphi_{\widetilde\Lambda}(y)$ in the inner loop algorithm from the expression in Proposition \ref{NP_thm_KKT_form} to the switching form
\[
    \tilde\varphi_{\widetilde\Lambda}(y)=\mathbbm{1}(D(y)>\delta_{SP})\varphi_S(y)+\mathbbm{1}(D(y)\leq\delta_{SP})\mathbbm{1}\left(g(y) \geq  \sum_{i=1}^{M_0} \tilde\lambda_i f_{\tilde\theta_i}(y)\right).
\]
See \cite{EMW15} for a formal definition of the standard best test $\varphi_S$ and further details on switching tests.


\subsubsection{Numerical Approximation Thresholds}

Acknowledging that we cannot perfectly compute $\bar\Omega^*$ (due to the numerical approximation of the outer loop) or $\varphi_{\Omega}^*$ for any $\Omega\in\Delta_{M_1}$ (due to the numerical approximation of the inner loop), let $\hat\varphi_{\widehat\Omega}^*$ denote the test obtained from the outer loop algorithm.  In addition to potential numerical error arising from the use of our algorithms, we must rely on approximations of rejection probabilities within step 2.(a) of the inner loop algorithm and step 2.(a) of the outer loop algorithm.  Nevertheless, the theoretical results in the following section and standard laws of large numbers imply that if we apply our algorithms with enough iterations and simulate rejection probabilities from enough Monte Carlo replications, these approximation errors will be small so that (i) $\int\hat\varphi_{\widehat\Omega}^*f_{\tilde\theta}d\nu$ cannot lie substantially above $\alpha$ for any $\tilde\theta\in\bar\Theta_0$ and (ii) $\int(\hat\varphi_{\widehat\Omega}^*-\varphi_{ah})f_{\theta}d\nu$ cannot lie substantially below zero for any $\theta\in\bar\Theta_1$.  We therefore use the following numerical approximation threshold rules for some small threshold $\epsilon>0$:
\begin{itemize}
\item For $\theta\in\Theta_0$, we conclude that $\int\hat\varphi_{\widehat\Omega}^*f_{\theta}d\nu\lessapprox\alpha$ if $\int\hat\varphi_{\widehat\Omega}^*f_{\theta}d\nu\leq\alpha+\epsilon$.
\item For $\theta\in\Theta_1$, we conclude that
\begin{itemize}
\item $\int(\hat\varphi_{\widehat\Omega}^*-\varphi_{ah})f_{\theta}d\nu \lessapprox    0$ if $\int(\hat\varphi_{\widehat\Omega}^*-\varphi_{ah})f_{\theta}d\nu < \varepsilon$,
\item $\int(\hat\varphi_{\widehat\Omega}^*-\varphi_{ah})f_{\theta}d\nu \gtrapprox  0$ if $\int(\hat\varphi_{\widehat\Omega}^*-\varphi_{ah})f_{\theta}d\nu > - \varepsilon$,
\item $\int(\hat\varphi_{\widehat\Omega}^*-\varphi_{ah})f_{\theta}d\nu \approx   0$ if $|\int(\hat\varphi_{\widehat\Omega}^*-\varphi_{ah})f_{\theta}d\nu | \leq \varepsilon$.
\end{itemize}
\end{itemize}

\subsubsection{Constructing the Approximate Power Envelope} \label{sec:add:supp:pts}

Using the numerical approximation threshold rules above, we implement the algorithm by potentially adding support points according to the following steps:
\begin{enumerate}
\item For a given $\bar\Theta_0$ and $\bar\Theta_1$, run the inner and outer loop algorithms until $\int\hat\varphi_{\widehat\Omega}^*f_{\tilde\theta}d\nu\lessapprox\alpha$ for all $\tilde\theta\in\bar\Theta_0$ and $\int(\hat\varphi_{\widehat\Omega}^*-\varphi_{ah})f_{\theta}d\nu\gtrapprox 0$ for all $\theta\in\bar\Theta_1$. Proceed to step 2.
\item For a fine grid $\Theta_0^f\subset\Theta_0$, compute $\int\hat\varphi_{\widehat\Omega}^*f_{\tilde\theta}d\nu$ for each $\tilde\theta\in\Theta_0^f$.
\begin{itemize}
    \item If $\int\hat\varphi_{\widehat\Omega}^*f_{\tilde\theta}d\nu\lessapprox\alpha$ for all $\tilde\theta\in\Theta_0^f$, proceed to step 3.
    \item Otherwise, add (some of) the values in $\Theta_0^f$ for which $\int\hat\varphi_{\widehat\Omega}^*f_{\tilde\theta}d\nu\not\lessapprox\alpha$ to $\bar\Theta_0$ and return to step 1.
\end{itemize}  
\item For a fine grid $\Theta_1^f\subset\Theta_1$, compute $\int(\hat\varphi_{\widehat\Omega}^*-\varphi_{ah})f_{\theta}d\nu$ for each $\theta\in\Theta_1^f$.
\begin{itemize}
\item If $\int(\hat\varphi_{\widehat\Omega}^*-\varphi_{ah})f_{\theta}d\nu\gtrapprox 0$ for all  $\theta\in\Theta_1^f$, then $\hat\varphi_{\widehat\Omega}^*$ constitutes an \textit{approximate power envelope} and either
\begin{itemize}
\item $\int(\hat\varphi_{\widehat\Omega}^*-\varphi_{ah})f_{\theta}d\nu\approx 0$ for all $\theta\in\Theta_1^f$ $\Rightarrow$ Conclude that $\varphi_{ah}$ is \textit{effectively} optimal.
\item $\int(\hat\varphi_{\widehat\Omega}^*-\varphi_{ah})f_{\theta}d\nu \not \lessapprox 0$ for some $\theta\in\Theta_1^f$ $\Rightarrow$ Conclude that $\varphi_{ah}$ is \textit{effectively} dominated.
\end{itemize}
\item Otherwise, add (some of) the values in $\Theta_1^f$ for which $\int(\hat\varphi_{\widehat\Omega}^*-\varphi_{ah})f_{\theta}d\nu \not \gtrapprox 0$ to $\bar\Theta_1$ and return to step 1.
\end{itemize}
\end{enumerate}

The motivation for step 2.~is to ensure $\hat\varphi_{\widehat\Omega}^*$ has approximately correct size, in analogy with step 8.~of \citeauthor{EMW15}'s (\citeyear{EMW15}) algorithm.  Similarly, step 3.~is used to ensure that the power function of $\hat\varphi_{\widehat\Omega}^*$ provides a good approximation to that of a WAP-maximizing test that yields a power envelope for $\varphi_{ah}$.

\subsubsection{Approximating Rejection Probabilities}

To approximate rejection probabilities, we rely on Monte Carlo simulation but incorporate several modifications designed to improve the numerical stability and convergence of our algorithms. These adjustments impose qualitative features that the true rejection probabilities are known to satisfy, thereby reducing simulation noise that would otherwise distort the optimization.

First, in all examples we consider, the power function of the ad hoc test is continuous in $\theta$. Independent simulation draws across different $\theta$ values produce artificially jagged power functions. To avoid this, we employ common random numbers by generating a single set of baseline draws and obtaining simulated values of $Y$ for each parameter value by transforming these baseline draws (i.e., by translating them by~$\theta$). This induces smoothness in the simulated rejection probabilities.

Second, we ensure that key moments of the distribution of $Y$ are matched exactly in the simulations. For example, when $Y$ is standard bivariate normal, we standardize the baseline draws to have mean zero and identity covariance. In some instances we also impose symmetry by symmetrizing the baseline draws. For the bivariate normal case, this guarantees that the marginal distributions are symmetric around zero. In the boundary-robust testing application of Section \ref{sec:Cox:app}, this ensures that the estimated power function of the two-sided $t$-test is symmetric, as implied by theory.

Third, we choose the random seed so that the estimated null rejection probabilities are close to the nominal level whenever the true null rejection probabilities are known to equal the nominal level. For example, in the linear IV model the CLR test is  similar by construction. If the simulation draws happen to make the CLR test appear to overreject at some points in $\bar\Theta_0$, then the algorithm---which searches for a WAP-maximizing test that weakly dominates the CLR test while satisfying the size constraint---may struggle to converge. Ensuring that simulated null rejection rates are close to their theoretical values avoids these problems and yields more reliable numerical results.

\section{Theoretical Justification of Numerical Implementation}\label{sec:convergence theory}

In this section, we present the theoretical justification for both the inner and outer loop algorithms for computing approximate WAP-maximizing tests that can produce a power envelope for a given ad hoc test.

\subsection{Inner Loop}

We now present the theoretical result guaranteeing the convergence of the inner loop algorithm for minimizing $\tilde\phi$ in \eqref{eq:reformulated dual problem}, which as discussed in Section \ref{sec:inner loop alg}, also solves the primal problem \eqref{eq:discrete null problem}.  This algorithm is a dual (projected) subgradient descent algorithm and therefore its convergence properties readily follow from known results in the literature.  Let 
        \begin{align*}
            \tilde\phi^{(k)} = \min_{i = 0, 1,\dots, k} \tilde\phi(\widetilde\Lambda^{(i)}).
        \end{align*}

\begin{theorem}\label{EMW_convergence}
    The inner loop algorithm described in Section \ref{sec:inner loop alg} satisfies for any $k\ge 1$,\footnote{As there are potentially multiple solutions $\widetilde\Lambda^\ast$ of the dual problem, the distance on the right hand side can be interpreted as the smallest distance of $\Lambda^{(0)}$ to the set of solutions.}
    \begin{align*}
        \tilde\phi^{(k)} - \min_{\widetilde\Lambda \ge 0} \tilde\phi(\widetilde\Lambda) \le \sqrt{M_0 \max\{1-\alpha, \alpha\}}\frac{\lVert \widetilde\Lambda^{(0)} - \widetilde\Lambda^\ast \rVert _2^2 + \sum_{i=0}^k \tilde h_i^2}{2\sum_{i=0}^k \tilde h_i} =: \widetilde\Gamma_k.
    \end{align*}
    The tuning parameter choice
    \begin{align*}
        \tilde h_i = \frac{\varepsilon}{\sqrt{M_0 \max\{1-\alpha, \alpha\}}}, \qquad \text{for } i = 0, 1, \ldots
    \end{align*}
    with $\varepsilon > 0$ implies
    \begin{align*}
        \tilde\phi^{(N)} - \min_{\widetilde\Lambda \ge 0} \tilde\phi(\widetilde\Lambda) \le \varepsilon
    \end{align*}
    for all $N \ge \max\{1-\alpha, \alpha\} M_0 \lVert \widetilde\Lambda^{(0)} - \widetilde\Lambda^\ast\rVert _2^2/\varepsilon^2$.
\end{theorem}

The convergence rate is controlled by $\widetilde\Gamma_k$, which depends on the starting value $\widetilde\Lambda^{(0)}$ through $\lVert \widetilde\Lambda^{(0)} - \widetilde\Lambda^\ast\rVert_2^2$ and on the tuning parameter sequence $\{\tilde h_k\}$. It holds that $\widetilde\Gamma_k \to 0$ as $k\to \infty$ if $\sum \tilde h_k = \infty$ and $\sum \tilde h_k^2 < \infty$.
The particular proposed choice of the tuning parameters $\{\tilde h_k\}$ is taken from Section 3.2.3 of \cite{nesterov2018lecture}. This choice is convenient as it only depends on known quantities and guarantees that the algorithm finds an $\varepsilon$-solution if the number of iterations is sufficiently large. 

\cite{EMW15} also propose an algorithm for approximating $\tilde\varphi_{\widetilde\Lambda^\ast}$ but they do not provide any convergence guarantees analogous to our Theorem \ref{EMW_convergence} for our inner loop algorithm.  However, while we were writing this paper, \cite{AFBMOQST25} proposed an algorithm that slightly modifies that of \cite{EMW15} and produces a test that provably approximates $\tilde\varphi_{\widetilde\Lambda^\ast}$ with convergence guarantees analogous to our Theorem \ref{EMW_convergence} as well as bounds the Monte Carlo error.  Their results also enable them to bound the size-distortions of their approximately optimal test and provide specific theoretically-grounded tuning parameter recommendations for an updating parameter, number of iterations and initial weights.  Either the original algorithm of \cite{EMW15}, the new modified algorithm of \cite{AFBMOQST25} or the linear-programming algorithm of \cite{MM13} can be used in place of our inner loop algorithm if desired.

\subsection{Outer Loop}\label{sec:outer loop conv}

Moving to the theoretical result guaranteeing the convergence of the outer loop algorithm for approximating the weight function that solves \eqref{eq:discrete power optimization}, we again note that since this algorithm is a (projected) subgradient descent algorithm, its convergence properties readily follow from known results in the literature.  Let 
\begin{align*}
            \phi^{(k)} = \min_{i = 0, 1,\dots, k} \phi(\bar\Omega^{(i)}).
        \end{align*}

\begin{theorem}\label{outer_loop_convergence}
    The outer loop algorithm described in Section \ref{sec:outer loop alg} satisfies for any $k\ge 1$,\footnote{As there are potentially multiple solutions $\bar\Omega^\ast$ of \eqref{eq:discrete power optimization}, the distance on the right hand side can be interpreted as the smallest distance of $\bar\Omega^{(0)}$ to the set of solutions.}
    \begin{align*}
        \phi^{(k)} - \min_{\bar\Omega \in \Delta_{M_1}} \phi(\bar\Omega) \le \sqrt{M_1}\frac{\lVert \Omega^{(0)} - \bar\Omega^\ast \rVert _2^2 + \sum_{i=0}^k h_i^2}{2\sum_{i=0}^k h_i} =: \Gamma_k.
    \end{align*}
    The tuning parameter choice 
    \begin{align*}
    h_i = \frac{\varepsilon}{\sqrt{M_1}}, \qquad \text{for } i = 0, 1, \dots
\end{align*}
with $\varepsilon>0$ implies
\begin{align*}
    \phi^{(N)} - \min_{\bar\Omega \in \Delta_{M_1}} \phi(\Omega) \le \varepsilon
\end{align*}
for all $N \ge M_1 \lVert \Omega^{(0)} - \bar\Omega^\ast\rVert _2^2/\varepsilon^2$.
\end{theorem}

The convergence rate is controlled by $\Gamma_k$, which depends on the starting value $\Omega^{(0)}$ through $\lVert \Omega^{(0)} - \bar\Omega^\ast\rVert_2^2$ and on the tuning parameter sequence $\{h_k\}$. It holds $\Gamma_k \to 0$ as $k\to \infty$ if $\sum h_k = \infty$ and $\sum h_k^2 < \infty$, which as with the inner loop is a convenient choice taken from Section 3.2.3 of \cite{nesterov2018lecture}.

\section{Applications} \label{sec:applications}
In this section, we apply our results and algorithms to shed new light on the optimality properties of a test whose optimality properties have already been thoroughly analyzed in the literature as well as a brand-new test that has not yet.  Specifically, we analyze the optimality of the CLR test of Moreira (2003) and the test implied by the inequality-imposed confidence interval of \cite{Cox24}. Throughout this section, the nominal level $\alpha$ is taken equal to 5\%.

\subsection{CLR Test in the Homoskedastic Linear IV Model} \label{sec:linearIV:app}
As alluded to in the introduction, \cite{AMS06,AMS08} (AMS06 and AMS08, henceforth) investigate the optimality of the CLR test in the homoskedastic linear IV model, holding the variance matrix for the reduced-form errors fixed in their analysis (coined as the ``fixed-$\Omega$ design'' by \citealp{VdSW23}). In this setting, AMS06 construct an asymptotically efficient two-sided point-wise power envelope for invariant similar tests as the rejection probabilities arising from a collection of point-optimal invariant similar tests, finding the striking result that the CLR test numerically attains this power envelope.  AMS08 further strengthen this result by showing that one essentially obtains the same power envelope without imposing similarity.  However, \cite{AMY19} (AMY, henceforth) provide a counterpoint to these optimality results by finding that the CLR test falls short of a point-wise power envelope that is constructed by varying the value of the parameter of interest and keeping its hypothesized value fixed, rather than the more standard analysis that keeps the value of the parameter of interest fixed and varies its hypothesized value.  In recent work, \cite{VdSW23} show that this latter analysis is essentially the same as the standard analysis that varies the parameter of interest while keeping its hypothesized value fixed but instead of holding the variance matrix for the reduced-form errors fixed, fixes the variance matrix of the structural and first-stage errors (coined as the ``fixed-$\Sigma$ design'' by \citealp{VdSW23}).  \cite{VdSW23} further argue that the fixed-$\Omega$ design implicitly favors the power function of the CLR test and that the fixed-$\Sigma$ design is better suited for analyzing power in cases of low to moderate endogeneity as well as differing signs of the parameter of interest and the correlation between the structural errors in the IV model.

These recent results of AMY and \cite{VdSW23} motivate us to revisit the optimality analysis of the CLR test, looking at both the fixed-$\Omega$ and fixed-$\Sigma$ designs.  Analyzing the CLR test under the fixed-$\Sigma$ design using our new numerical optimality assessment enables us to assess whether the previously examined point-wise power envelopes of AMY may simply be setting an unattainable power bound.  Before proceeding, we formally introduce the linear IV model and the CLR test, and briefly describe the power envelope of AMS06. We closely follow the notation of AMS06. 

The linear IV model is given by
\begin{align*}
	y_1 &= y_2 \beta + u,\\
	y_2 &= Z \pi + v_2,
\end{align*}
where $y_1,y_2, u,v_2 \in \mathbbm{R}^n$, $Z \in \mathbbm{R}^{n\times k}$, $\beta \in \mathbbm{R}$, and $\pi \in \mathbbm{R}^k$. Here, $y_1$, $y_2$ and $Z$ are observed, where $y_1$ denotes the outcome of interest, $y_2$ the (potentially) endogenous regressor of interest, and $Z$ the instruments.\footnote{We omit additional exogenous regressors without loss of generality, as the above model can always be obtained by partialling them out.} The random vectors $u$ and $v_2$ are unobserved structural error terms. We assume that $Z$ is fixed. Plugging the reduced-form equation for $y_2$ into the structural equation for $y_1$, we obtain the reduced-form equation for $y_1$, i.e.,
\[
	y_1 = Z \pi \beta + v_1,
\]
where $v_1 = u + v_2\beta$. The resulting set of reduced-form equations can be written as\footnote{Here, we use $Y$ to be consistent with the notation in AMS06. This $Y$ should, however, not be confused with the random element $Y$ that enters the general testing problem in Section \ref{sec: testing problem}.}
\[
	Y = Z\pi a' + V, 
\]
where
\[
	Y = [y_1, y_2], \ V = [v_1, v_2] \text{ and } a = (\beta,1)'.
\]
Let $V_i$ denote the $i^\text{th}$ row of $V$ and assume that $V_i$ is iid across $i$ with 
\[  
    V_i \sim \mathcal{N}(0,\Omega).
\]
Having defined the linear IV model, we can now formally state the testing problem of interest, which is given by
\[
    H_0: \ \beta = \beta_0, \ \pi \in \mathbbm{R}^k \text{ vs.\ } H_1: \  \beta \in \mathbbm{R}\backslash \{ \beta_0 \}, \ \pi \in \mathbbm{R}^k.
\]
We follow AMS06 and further simplify this testing problem. To that end,  we define the following transformations of $Y$:
\[
	S = (Z'Z)^{-1/2}Z'Yb_0\cdot(b_0'\Omega b_0)^{-1/2} \text{ and }
	T = (Z'Z)^{-1/2}Z'Y \Omega^{-1}a_0 \cdot (a_0'\Omega a_0)^{-1/2},
\]
where $b_0 = (1,-\beta_0)'$ and $a = (\beta_0,1)'$. Lemma 2 of AMS06 shows that $S$ and $T$ are jointly normally distributed and independent. AMS06 argue that the coordinate system used to specify $S$ and $T$ should not affect inference and, therefore, only consider statistics that are invariant to rotation of the coordinate system. This is achieved by  considering test statistics that only depend on $S$ and $T$ through 
\[	
	Q = \left[ \begin{array}{cc} Q_S & Q_{ST} \\ Q_{ST} & Q_T \end{array} \right]
	= \left[ \begin{array}{cc} S'S & S'T \\ T'S & T'T \end{array} \right];
\]
see Theorem 1 in AMS06 and the surrounding discussion. The distribution of $Q$ is noncentral Wishart and, importantly, depends on $\pi$ only through
\[
    \lambda = \pi'Z'Z\pi,
\]
which implies that if we restrict our attention to test statistics based on $Q$, the testing problem of interest simplifies to
\begin{equation} \label{eq:H0:linearIV}
    H_0: \ \beta = \beta_0, \ \lambda \in \mathbbm{R}_+ \text{ vs.\ } H_1: \  \beta \in \mathbbm{R}\backslash \{ \beta_0 \}, \ \lambda \in \mathbbm{R}_+.
\end{equation}
The LR statistic for testing \eqref{eq:H0:linearIV} can be written as 
\[
	\text{LR} = \frac{1}{2} \left( Q_S - Q_T + \sqrt{(Q_S - Q_T)^2 + 4 Q_{ST}^2} \right).
\]
Moreira (2003) observes that $Q_T$ is a sufficient statistic for $\lambda$ under $H_0$. This, in turn, allows the construction of a size $\alpha$ test using \textit{conditional} critical values. In particular, the CLR test rejects when $\text{LR} > \text{cv}_{1-\alpha}(Q_T)$, where $\text{cv}_{1-\alpha}(Q_T)$ is such that $P_{\beta_0}(\text{LR} > \text{cv}_{1-\alpha}(Q_T)|Q_T) = \alpha$.


The power envelope proposed by AMS06 is constructed from a collection of point-optimal invariant similar two-sided (POIS2) tests. Invariance is imposed by restricting attention to tests that only depend on the data through $Q$. In the context at hand, imposing similarity is tantamount to relying on conditional critical values, conditional on the observed value of $Q_T$. This avoids the need for approximating the least-favorable distribution, as done in AMS08. The two-sidedness that AMS06 consider is such that the resulting test is asymptotically efficient, i.e., the test has the same power as the two-sided $t$-test when instruments are strong ($\lambda = \infty$). The corresponding point-optimal test, POIS2, puts equal weight on $(\beta^*,\lambda^*)$ and $(\beta^*_2,\lambda^*_2)$, where the latter is a function of the former, ensuring asymptotic efficiency. The corresponding power envelope is then mapped out by the POIS2 test as $(\beta^*,\lambda^*)$ varies.

\subsubsection{Fixed-$\Omega$ Design}
\begin{figure}[h]
  \centering
  \begin{subfigure}[b]{0.495\textwidth}
    \centering
    \includegraphics[width=\textwidth]{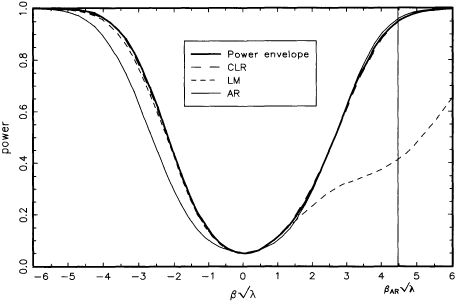}
    \caption{Screenshot of Figure 1(c) in AMS06}
  \end{subfigure}
  \hfill
  \begin{subfigure}[b]{0.495\textwidth}
    \centering
    \includegraphics[width=\textwidth]{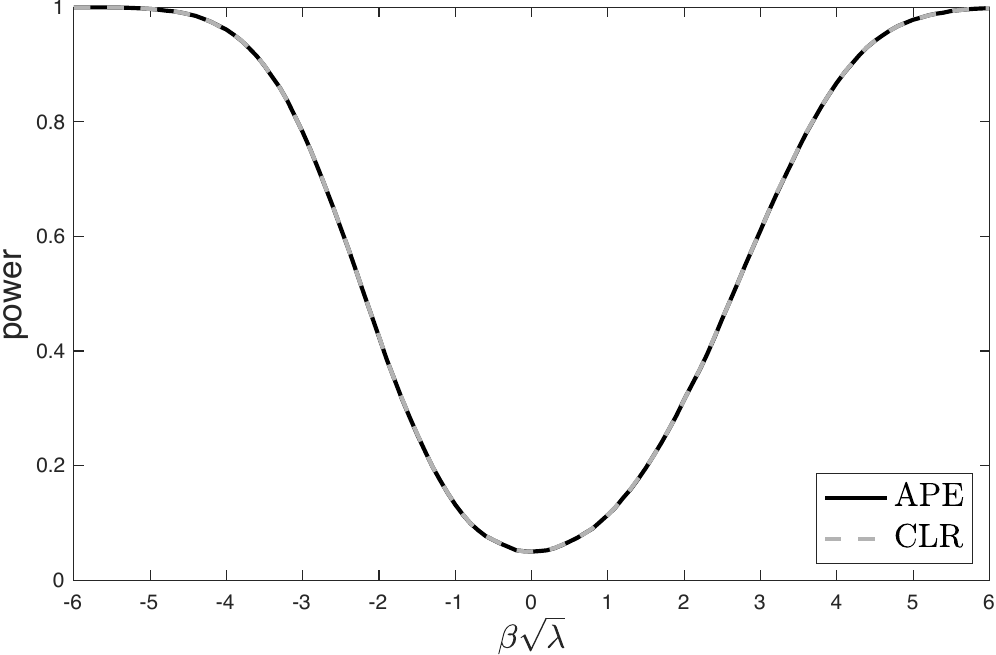}
    \caption{Power of CLR test together with our APE}
  \end{subfigure}
  \caption{Power of CLR test as a function of $\beta$ for $\lambda = 5$ together with AMS06's power envelope and our approximate power envelope for $k = 5$ and $\Omega_{12} = 0.5$.}
  \label{Figure:AMS06F1c}
\end{figure}

AMS06 consider a fixed-$\Omega$ design in their simulations, setting $\Omega_{11} = \Omega_{22} = 1$ and $\beta_0 = 0$. They numerically compare the power of the CLR test to their point-wise power envelope as functions of $\beta$ and $\lambda$ for various values of $k$ and $\Omega_{12}$ ($\rho$ in their paper).  For given values of $k$ and $\Omega_{12}$, they consider several $\lambda$-slices of the CLR power function and the power envelope, for which $\beta$ is varied over a fine grid for a given value of $\lambda$. An example of such a $\lambda$-slice is given in Figure \ref{Figure:AMS06F1c}.

Panel (a) of Figure \ref{Figure:AMS06F1c} is a screenshot of Figure 1(c) in AMS06, which shows the power of the CLR test (and the Anderson-Rubin and Lagrange multiplier tests of \citealp{AR49} and \citealp{Kle02}) together with their proposed power envelope as $\beta$ varies and $\lambda = 5$ for $k = 5$ and $\Omega_{12} = 0.5$. The power of the CLR test and the power envelope are virtually indistinguishable (or ``very close'' in this and many other figures in AMS06 where $k$, $\Omega_{12}$, and $\lambda$ are varied), which leads AMS06 to conclude that the CLR test attains the power envelope ``in a numerical sense''.

We reconsider the above testing problem for $k = 5$ and $\Omega_{12} = 0.5$. In the case at hand, $Q$ takes the role of $Y$ and $\lambda$ replaces $\delta$.\footnote{In principle, we could use $(S,T)$ as $Y$, replace $\delta$ by $\pi$ and investigate the optimality of the CLR test in the class of asymptotically efficient tests. However, the computational cost of approximating of $\Lambda$ increases rapidly in the dimension of the nuisance parameter.} We set $\lambda_\text{S} = 75$ and $\lambda_\text{SP} = 160$, where $Q_T$ takes the role of $D(Y)$, since the probability that $Q_T>\lambda_{SP}$ is less than $0.01$ whenever $\lambda\leq\lambda_S$. The standard test is implemented as the Lagrange multiplier test, which rejects $H_0$ when $Q_{ST}^2/Q_T > \chi^2_{1-\alpha}(1)$.\footnote{For $\lambda > \lambda_\text{S} = 75$, the CLR test, the Lagrange multiplier test and the two-sided $t$-test all approximately coincide.} Note that switching to the standard test (for large $\lambda$) is equivalent to imposing asymptotic efficiency as defined by AMS06. We obtain our APE using $\bar \Theta_0 = \{ (\beta,\lambda) : \beta = 0,\ \lambda \in \{1,5,10,\dots,30,40,\dots,170\} \}$ and $\bar \Theta_1 = \{ (b/\sqrt{\lambda},\lambda) : b\in \{-4,-3,-2,2,3,4\},\ \lambda \in \{ 1,5,10,\dots,30,40,\dots,170 \} \}$. Implementation details, including the number of simulation draws and the choices of $\{h_k\}$, $\Theta_0^f$ and $\Theta_1^f$ are given in Appendix \ref{app:details:applications}.

\begin{figure}[h]
    \centering
    \includegraphics[width=0.7\textwidth]{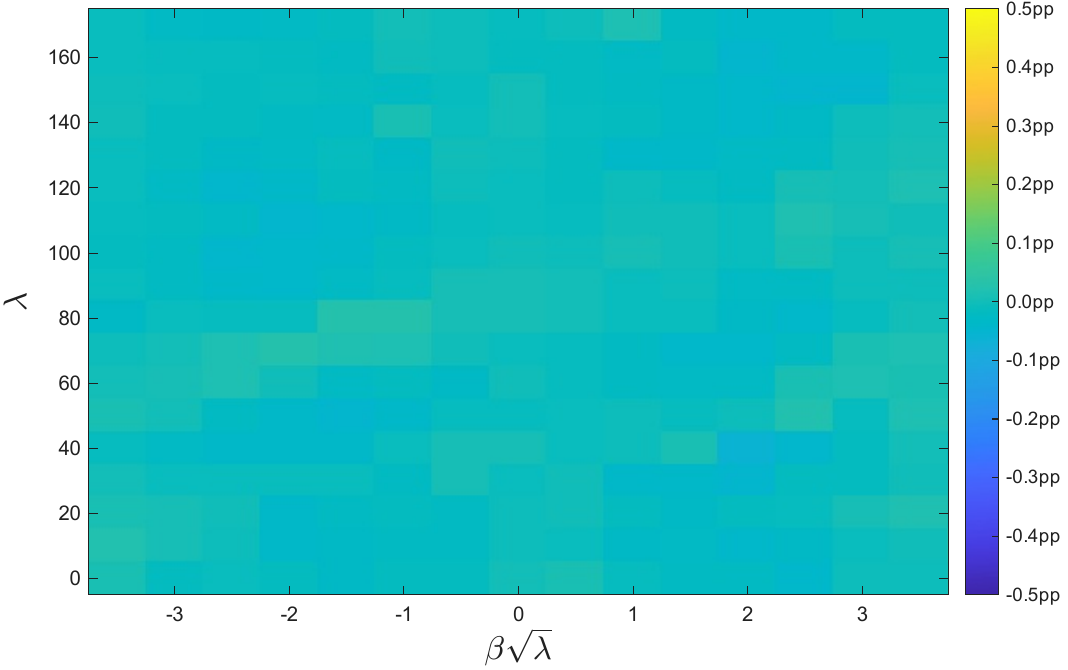}
  \caption{Difference between APE and power of CLR test in percentage points for $k = 5$ and $\Omega_{12} = 0.5$.}
  \label{Figure:AMS06F1c:heatmap}
\end{figure}

Panel (b) of Figure \ref{Figure:AMS06F1c} shows the power function of the CLR test (for the same parameter constellation as in panel (a)) together with our APE. As in panel (a), the power of the CLR test is indistinguishable from the power envelope for the $\lambda$-slice under consideration. Instead of considering multiple $\lambda$-slices, we produce a heatmap showing the difference between our APE and the power of the CLR test over a grid of values for $\beta$ and $\lambda$.\footnote{\label{fn:wc}The grid is given by $\{ (b/\sqrt{\lambda},\lambda) : b\in \{-3.5,-3,\dots,3.5\},\ \lambda \in \{ 0.1,10,20,\dots,170 \} \}$. Note that this grid does not coincide with $\bar \Theta_0 \cup \bar{\Theta}_1$ and that the underlying rejection probabilities are evaluated using draws that are independent from those used to obtain the APE. This protects from a potential winner's curse, cf.~\cite{AKM24}.} This heatmap is shown in Figure \ref{Figure:AMS06F1c:heatmap}.

The differences between the APE and the power of the CLR test are very close to zero over the entire grid; note that the scale on the right ranges only from -0.5 percentage points (pp) to 0.5pp. In fact, the largest difference (in absolute value) is below 0.1pp. We therefore conclude that the CLR test is effectively optimal (for $k = 5$ and $\Omega_{12} = 0.5$). Our finding is complementary to that of AMS06:~we find that the CLR test is (effectively) optimal in the class of invariant asymptotically efficient tests, while AMS06 find the stronger result of point-optimality in the smaller class of tests that are similar and two-sided (in the sense that they impose).\footnote{Our finding is also complementary to that of AMS08:~while we find a weaker result (optimality as opposed to point-optimality), we do not impose two-sidedness (in the sense that they impose).}

\subsubsection{Fixed-$\Sigma$ Design}
As shown by \cite{VdSW23}, AMY consider a fixed-$\Sigma$ design in their simulations, where the variance matrix of the {structural} errors $u$ and $v_2$, $\Sigma$, is held constant. AMY set $\Sigma_{11} = \Sigma_{22} = 1$ and consider several values for $\Sigma_{12}$ ($\rho_{uv}$ in their paper). Their Figure 1 shows the power functions of the CLR test (and the Anderson-Rubin test) and the power envelope of AMS06 based on the POIS2 test for different values of $\Sigma_{12}$, keeping $k=10$ and $\lambda = 15$ fixed. In contrast to AMS06, however, AMY set $\beta = 0$ and vary $\beta_0$.

\begin{figure}[h]
  \centering
  \begin{subfigure}[b]{0.495\textwidth}
    \centering
    \includegraphics[width=\textwidth]{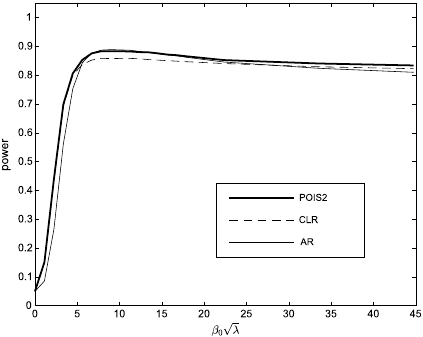}
    \caption{Screenshot of Figure 1(b) in AMY}
  \end{subfigure}
  \hfill
  \begin{subfigure}[b]{0.495\textwidth}
    \centering
    \includegraphics[width=\textwidth]{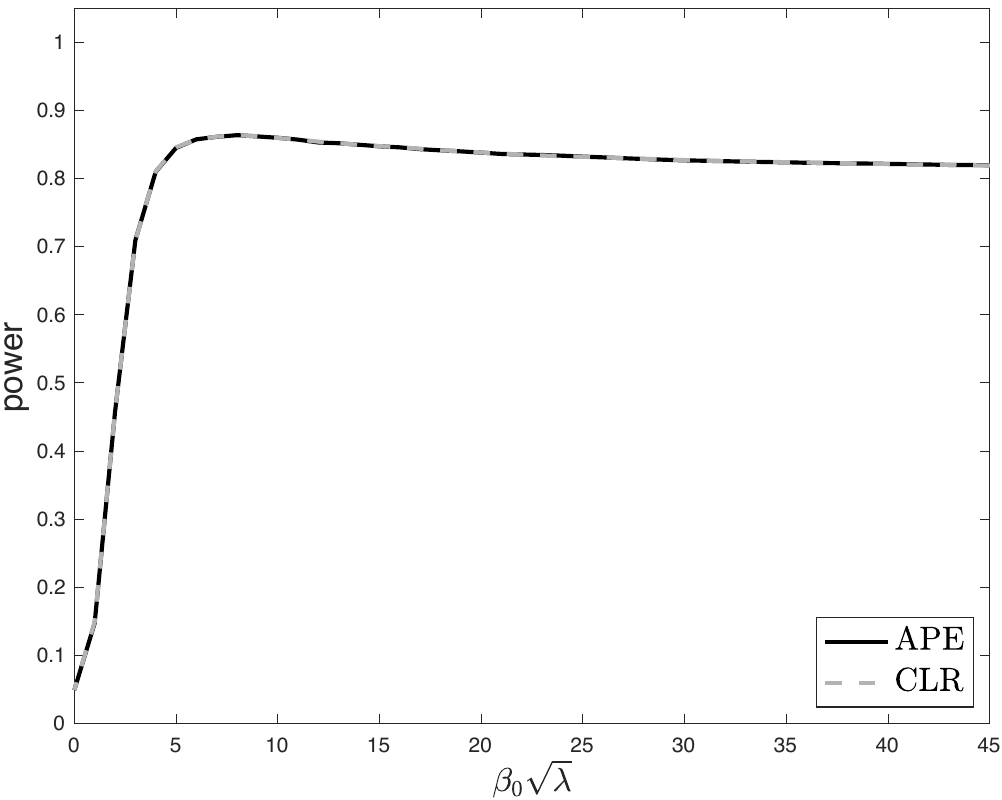}
    \caption{Power of CLR test together with our APE}
  \end{subfigure}
  \caption{Power of CLR test as a function of $\beta_0$ for $\lambda = 15$ together with AMS06's power envelope and our APE for $k = 10$ and $\Sigma_{12} = 0.5$.}
  \label{Figure:AMYF1b}
\end{figure}

Panel (a) of Figure \ref{Figure:AMYF1b} is a screenshot of Figure 1(b) in AMY, where $\Sigma_{12} = 0.5$. The power of the CLR test is on the power envelope for values of $\beta_0 \sqrt{\lambda}$ between 0 and 5, but drops below for values further away from 0. The maximal gap of the CLR test's power function with the POIS2 power envelope is around 3--4pp. Based on this finding, AMY conclude that the finding of AMS06 ``that the CLR test is essentially on the [...] power envelope does not hold...''. However, as discussed extensively above, this does not mean that the CLR test is not optimal since it is unclear if the point-wise power envelope used by AMY is attainable. We therefore reconsider the testing problem considered in AMY, where $k = 10$ and $\Sigma_{12} = 0.5$, and construct our APE. We set $\lambda_\text{S} = 160$ and $\lambda_\text{SP} = 320$. The underlying $\bar \Theta_0$ and $\bar \Theta_1$ as well as additional implementation details are provided in Appendix \ref{app:details:applications}.

\begin{figure}[h]
    \centering
    \includegraphics[width=0.7\textwidth]{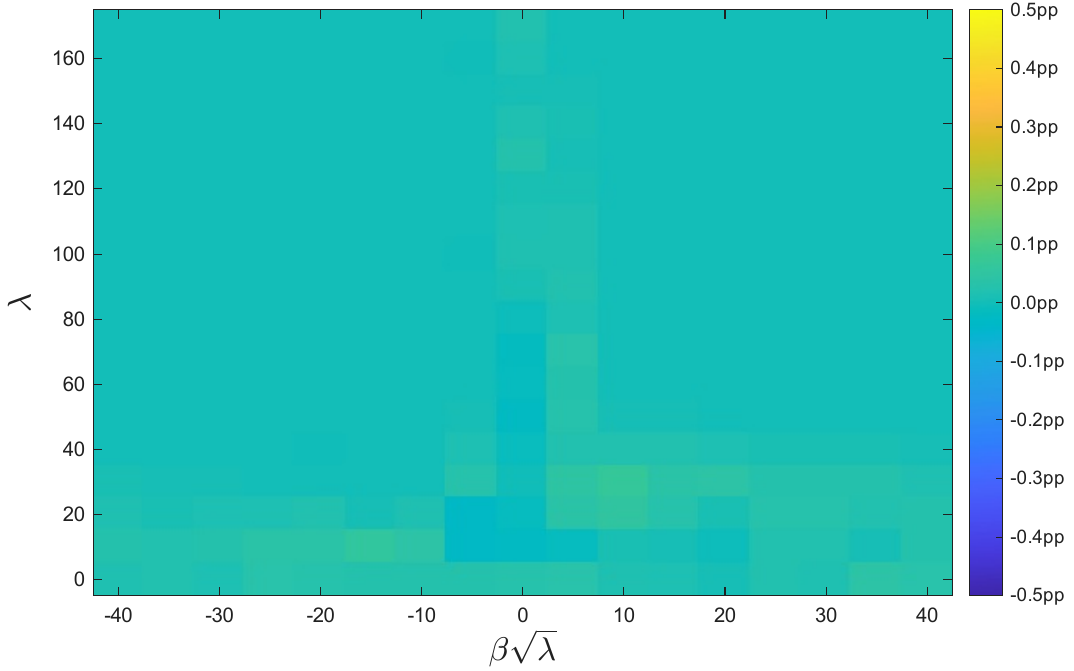}
  \caption{Difference between APE and power of CLR test in percentage points for $k = 10$ and $\Sigma_{12} = 0.5$.}
  \label{Figure:AMYF1b:heatmap}
\end{figure}

Panel (b) of Figure \ref{Figure:AMYF1b} displays the power function of the CLR test (for the same parameter constellation as in panel (a)) together with our APE. We find that the CLR test attains our APE, at least for the $\lambda$-slice under consideration. As before, we produce a heatmap showing the difference between our APE and the power of the CLR test over a grid of values for $\beta$ and $\lambda$, while setting $\beta_0 = 0$. While this setting seemingly differs from the setting of AMY (where $\beta = 0$ while $\beta_0$
varies), it follows from Corollary 1 of \cite{VdSW23} and the subsequent discussion that the resulting power curves are simply mirror images of one another.\footnote{That is, the power of the CLR test for testing $H_0: \beta = \beta_0$ when the true value of $\beta$, say $\beta^*$, is equal to 0, is equal to the power of the CLR test for testing $H_0: \beta = 0$ when $\beta^* =- \beta_0$.} The heatmap is shown in Figure \ref{Figure:AMYF1b:heatmap}.

The heatmap in Figure \ref{Figure:AMYF1b:heatmap} uses the same scale as the heatmap in Figure \ref{Figure:AMS06F1c:heatmap} and leads us to the same conclusion:~the difference between our APE and the power function of the CLR test is very close to zero across the entire grid. We conclude that the CLR test is, in fact, effectively optimal \emph{under both} the fixed-$\Omega$ and fixed-$\Sigma$ designs.


\subsection{Test Implied by IICI in Boundary Problem}\label{sec:Cox:app}
A line of recent work has examined testing problems involving uniformly-valid inference when nuisance parameters may lie on or near the boundary of the parameter space.\footnote{See, for example, \cite{AG10b}, \cite{Ket18}, \cite{KM25} and \cite{Cavaliere2025}.} Focusing on the special case of a scalar nuisance parameter to analyze the properties of a new confidence interval proposed by \cite{Cox24}, we work with the Gaussian experiment
\[ 
    Y = \left( \begin{array}{c} Y_1\\Y_2 \end{array} 
\right) \sim N\left( \left(\begin{array}{c} \beta \\ \delta \end{array}\right),\left(\begin{array}{cc} 1 & \rho \\ \rho & 1 \end{array} \right)\right),
\]
where $\beta$ is the scalar parameter of interest and $\delta \ge 0$ is a scalar nuisance parameter. 
As discussed in Section~\ref{sec: testing problem}, this formulation corresponds to the asymptotic behavior of a broad class of finite-sample models (see, e.g.,~\citealp{EMW15} and \citealp{KM25}). Formally, the testing problem of interest is given by\footnote{This limiting Gaussian formulation also applies to problems for which nuisance parameters are restricted to be greater/less than or equal to any known value via a simple affine transformation.}
\begin{equation} \label{eq:H0:bdy}
    H_0: \beta = \beta_0,\ \delta \geq 0  \text{ vs.\ } H_1: \beta \neq \beta_0, \ \delta \geq 0.
\end{equation}

\cite{Cox24} proposes a new confidence interval for $\beta$, called the inequality-imposed confidence interval (IICI). The IICI has several desirable properties:~(i) it is easy to compute, (ii) it does not require simulation or tuning parameters, (iii) it is adaptive, and (iv) it has weakly shorter length than the standard two-sided confidence interval that ignores the information contained in $Y_2$. The IICI is constructed as the union of the standard two-sided CI when $Y_2$ is large and positive, the two-sided confidence interval that imposes $\delta = 0$ when $Y_2$ is large and negative, and the intersection of the two when $Y_2$ is close to zero; see Panel A of Figure~1 in \cite{Cox24} for a visual illustration. Formally, the lower and upper bounds of the $(1-\alpha)$-nominal IICI are given by
\[
    \begin{cases}
        Y_1 - z_{1-\alpha/2} & \text{ if } Y_2 > c, \\
        Y_1 - \rho Y_2 - \sqrt{1-\rho^2}\, z_{1-\alpha/2} & \text{ otherwise,}
    \end{cases}
\]
and
\[
    \begin{cases}
        Y_1 + z_{1-\alpha/2} & \text{ if } Y_2 > -c, \\
        Y_1 - \rho Y_2 + \sqrt{1-\rho^2}\, z_{1-\alpha/2} & \text{ otherwise,}
    \end{cases}
\]
respectively, where $c = \frac{1 - \sqrt{1-\rho^2}}{\rho} z_{1-\alpha/2}$ and $z_{1-\alpha/2}$ denotes the $1-\alpha/2$ quantile of the $\mathcal{N}(0,1)$ distribution. The test implied by the IICI rejects $H_0$ in \eqref{eq:H0:bdy} when $\beta_0$ lies outside of it.

\begin{figure}[h]
    \centering
    \includegraphics[width=0.6\textwidth]{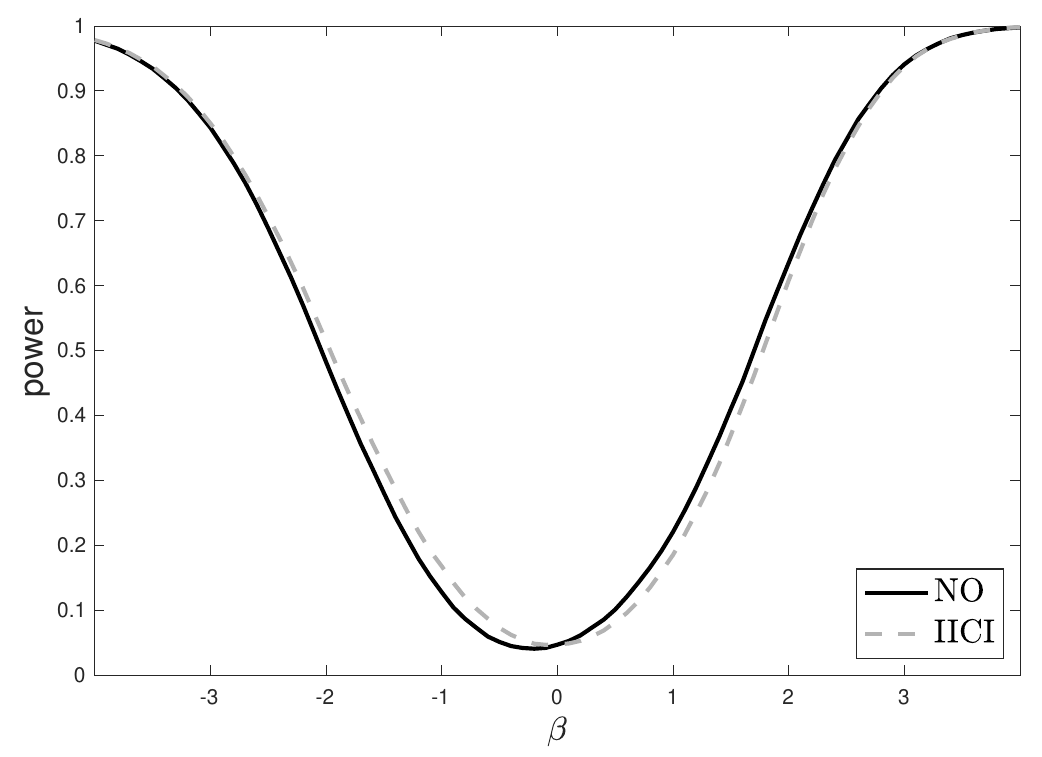}
  \caption{Power of the nearly optimal test of \cite{EMW15} and the test implied by the IICI for $\rho = 0.7$ and $\delta = 1$.}
  \label{Figure:IICI:VS:NO}
\end{figure}

To gauge the optimality of the IICI, \cite{Cox24} compares the WAP of the test implied by the IICI with the power bound on WAP derived in \cite{EMW15} (EMW, henceforth), using equal weight on $\beta=-2$ and $\beta=2$ and uniform weights on $\delta \in [0,9]$. For $\rho = 0.7$, the WAP of the IICI is 53.1\% and the WAP of the EMW power bound is 53.5\%. Using $\epsilon = 0.005$ (following EMW), \cite{Cox24} concludes that the test implied by the IICI is nearly optimal in the sense of EMW, as its WAP lies within $\epsilon$ of the power bound. However, some ambiguity with respect to the optimality of the test remains:\footnote{This is related to the fact that $\epsilon$ is given as an absolute (rather than a relative) value, making it difficult to interpret what it means for the difference in WAPs to be ``small''.} as shown in Figure~\ref{Figure:IICI:VS:NO}, the power functions of the test implied by the IICI and the nearly optimal test of EMW (under the above weights) cross, which prevents us from concluding whether the test implied by the IICI is optimal or dominated. 

\begin{figure}[h]
    \centering
    \includegraphics[width=0.6\textwidth]{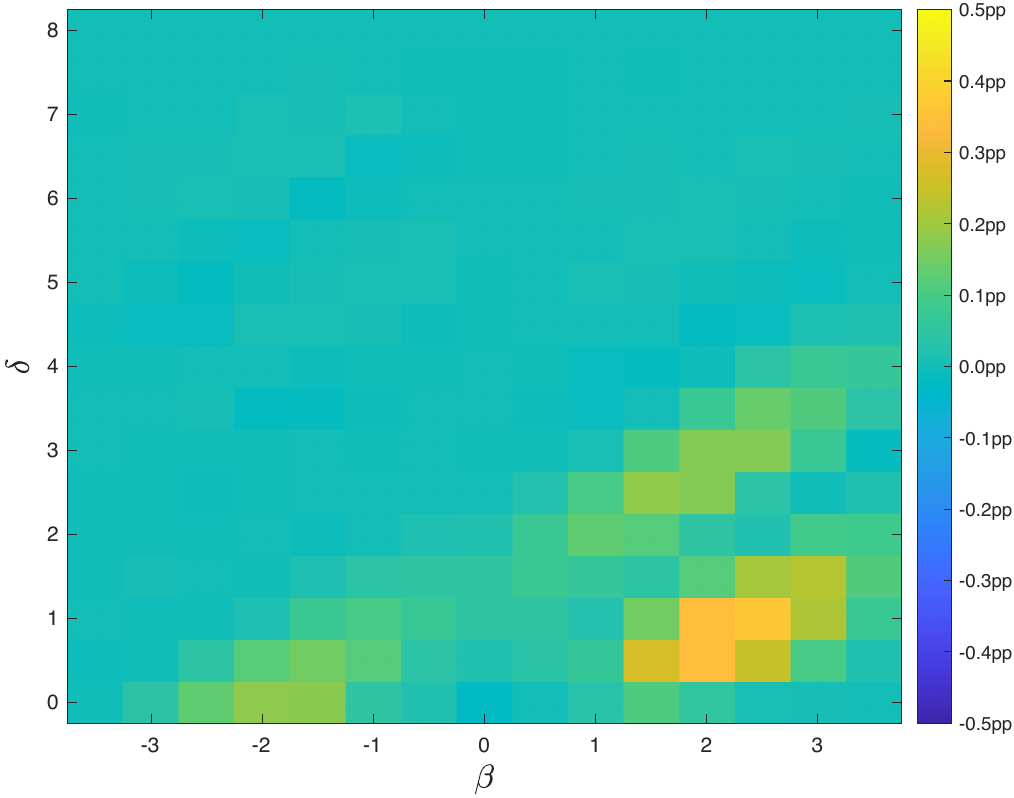}
  \caption{Difference between APE and power of test implied by the IICI in percentage points for $\rho = 0.7$.}
  \label{Figure:Cox:heatmap}
\end{figure}

To address this ambiguity, we compute our APE for the testing problem with $\rho = 0.7$. We follow EMW and choose $\delta_{\text{SP}} = 6$. The standard test is the two-sided $t$-test. We use $\bar{\Theta}_1 = \{ (b,d) : b \in \{-3,-2,-1,1,2,3\},\ d \in \{0,0.5,\dots,8\} \}$
and follow EMW in discretizing $\Theta_0$ in terms of ``base'' distributions. Details on this and other implementation choices are provided in Appendix~\ref{app:details:applications}. As in Section~\ref{sec:linearIV:app}, we produce a heatmap showing the difference between our APE and the power of the test implied by the IICI over a grid of values for $\beta$ and $\delta$, fixing $\beta_0 = 0$; see Figure~\ref{Figure:Cox:heatmap}.

The scale of the heatmap is the same as for the heatmaps in Section \ref{sec:linearIV:app}. The difference between the APE and the power of the test implied by the IICI is close to zero over a large portion of the grid. However, there are some values for which the test implied by the IICI falls short of the APE—the maximal difference is about 0.3pp and occurs at $(\beta,\delta) = (2,1)$. We, therefore, conclude that the test implied by the IICI is effectively dominated, albeit by a very small margin. In fact, the WAPs of the test underlying the APE and the test implied by the IICI are 52.532\% and 52.529\%, respectively, with the corresponding weights given in Appendix \ref{app:details:applications}. Abstracting from numerical approximations, the WAP of the test implied by the IICI is thus within 0.00003 of the lowest possible upper bound (given $\bar \Theta_1$), improving on the finding in \cite{Cox24} and providing a strong argument in favor of using the IICI in practice. Although the test is effectively dominated, the loss in WAP is negligible considering the simplicity of the procedure.


\newpage
\appendix

\section{Appendix}\label{sec:appendix}

\subsection{Proofs of Main Results}

\textbf{Proof of Theorem \ref{thm:WAP-max power bound equivalence}:} Starting with the first claim of the theorem, for any $\varphi \in \Phi_\alpha$, 
    \begin{align}
        \min_{j=1,\dots, M_1} \int (\varphi - \varphi_{ah}) f_{\theta_j} d\nu 
        = &\inf_{\bar\Omega \in \Delta_{M_1}} \sum_{j=1}^{M_1} \omega_j\int (\varphi - \varphi_{ah}) f_{\theta_j} d\nu. \label{eq:min over j or simplex}
    \end{align}
This can be seen as follows:~for any $\bar \Omega \in \Delta_{M_1}$,
    \begin{align*}
        \sum_{j=1}^{M_1} \omega_j\int (\varphi - \varphi_{ah}) f_{\theta_j} d\nu
        \ge \min_{j=1,\dots, M_1} \int (\varphi - \varphi_{ah}) f_{\theta_j} d\nu \sum_{i=1}^{M_1} \omega_i = \min_{j=1,\dots, M_1} \int (\varphi - \varphi_{ah}) f_{\theta_j} d\nu,
    \end{align*}
    and therefore
    \begin{align*}
        \inf_{\bar\Omega \in \Delta_{M_1}} \sum_{j=1}^{M_1} \omega_j\int (\varphi - \varphi_{ah}) f_{\theta_j} d\nu \ge \min_{j=1,\dots, M_1} \int (\varphi - \varphi_{ah}) f_{\theta_j} d\nu.
    \end{align*}
    Moreover, since the canonical basis vectors of $\mathbb{R}^{M_1}$ are included in $\Delta_{M_1}$, 
    \begin{align*}
        \min_{j=1,\dots, M_1} \int (\varphi - \varphi_{ah}) f_{\theta_j} d\nu
        \ge \inf_{\bar\Omega \in \Delta_{M_1}} \sum_{j=1}^{M_1} \omega_j\int (\varphi - \varphi_{ah}) f_{\theta_j} d\nu
    \end{align*}
    and \eqref{eq:min over j or simplex} follows. Thus, 
    \begin{align*}
        \sup_{\varphi \in \Phi_\alpha} \min_{j=1,\dots, M_1} \int (\varphi - \varphi_{ah}) f_{\theta_j} d\nu
        = &\sup_{\varphi \in \Phi_\alpha} \inf_{\bar\Omega \in \Delta_{M_1}} \sum_{j=1}^{M_1} \omega_j\int (\varphi - \varphi_{ah}) f_{\theta_j} d\nu.
    \end{align*}
    Further, in Lemma \ref{minimax_theorem}, we show that the Ky Fan minimax theorem implies
    \begin{align}
        \sup_{\varphi \in \Phi_\alpha} \inf_{\bar\Omega \in \Delta_{M_1}} \sum_{j=1}^{M_1} \omega_j\int (\varphi - \varphi_{ah}) f_{\theta_j} d\nu
        = &\inf_{\bar\Omega \in \Delta_{M_1}} \sup_{\varphi \in \Phi_\alpha} \sum_{j=1}^{M_1}\omega_j\int (\varphi - \varphi_{ah}) f_{\theta_j} d\nu.
        \label{conjecture_aux_2}
    \end{align}

The fact that $\Phi^*\neq \emptyset$ follows directly from Lemma \ref{existence_minimax_test}.
    
    To see that $\Delta_{M_1}^*\neq \emptyset$, note that since the criterion function on the right hand side of \eqref{conjecture_aux_2} is continuous in $(\bar\Omega, \varphi)$, $\phi : \Delta_{M_1} \to \mathbb{R}$ given by
    \begin{align*}
        \phi(\bar\Omega) = \sup_{\varphi \in \Phi_\alpha} \sum_{j=1}^{M_1}\omega_j\int (\varphi - \varphi_{ah}) f_{\theta_j} d\nu,
    \end{align*}
    is {lower} semi-continuous and therefore by compactness of $\Delta_{M_1}$, the infimum is attained at some $\bar\Omega^\ast = (\omega_1^\ast, \dots, \omega_{M_1}^\ast) \in \Delta_{M_1}$:
    \begin{align}
        \sup_{\varphi \in \Phi_\alpha} \sum_{j=1}^{M_1}\omega_j^\ast \int (\varphi - \varphi_{ah}) f_{\theta_j} d\nu
        = &\inf_{\bar\Omega \in \Delta_{M_1}} \sup_{\varphi \in \Phi_\alpha} \sum_{j=1}^{M_1}\omega_j\int (\varphi - \varphi_{ah}) f_{\theta_j} d\nu. 
        \label{conjecture_aux_3}
    \end{align}
    
    To show \eqref{eq:WAP:maximizer:is:maximin}, note that for any $\varphi^\ast \in \Phi^\ast$
    \begin{align*}
        \sum_{j=1}^{M_1} \omega_j^\ast \int(\varphi^\ast - \varphi_{ah}) f_{\theta_j} d\nu
        \ge &\inf_{\bar\Omega\in \Delta_{M_1}} \sum_{j=1}^{M_1} \omega_j \int(\varphi^\ast - \varphi_{ah}) f_{\theta_j} d\nu \\
        = &\sup_{\varphi\in \Phi_\alpha} \inf_{\bar\Omega\in \Delta_{M_1}} \sum_{j=1}^{M_1} \omega_j \int(\varphi - \varphi_{ah}) f_{\theta_j} d\nu\\
        = &\inf_{\bar\Omega\in \Delta_{M_1}} \sup_{\varphi\in \Phi_\alpha} \sum_{j=1}^{M_1} \omega_j \int(\varphi - \varphi_{ah}) f_{\theta_j} d\nu\\
        = &\sup_{\varphi\in \Phi_\alpha} \sum_{j=1}^{M_1} \omega_j^\ast \int(\varphi - \varphi_{ah}) f_{\theta_j} d\nu,
    \end{align*}
    where the inequality follows by definition of the infimum, the first equality by the definition of $\Phi^*$ and \eqref{eq:min over j or simplex}, the second by \eqref{conjecture_aux_2} and the last by \eqref{conjecture_aux_3}. By the definition of the supremum, we also have
    \[
        \sup_{\varphi\in \Phi_\alpha} \sum_{j=1}^{M_1} \omega_j^\ast \int(\varphi - \varphi_{ah}) f_{\theta_j} d\nu \geq
         \sum_{j=1}^{M_1} \omega_j^\ast \int(\varphi^\ast - \varphi_{ah}) f_{\theta_j} d\nu.
    \]
    Therefore, the inequality in the preceding display can be replaced by an equality and \eqref{eq:WAP:maximizer:is:maximin} follows. $\square$

\noindent
\textbf{Proof of Theorem \ref{thm:N-P test power envelope}:} 
Under the given conditions, the first statement follows directly from Lemma \ref{pot_power_fun_inner_loop_new}. The second statement follows from the first, together with equation \eqref{eq:WAP:maximizer:is:maximin}, which implies that $\varphi_1^\ast$ and $\varphi_2^\ast$ are WAP-maximizing tests with respect to $\bar\Omega_1^\ast$ and $\bar\Omega_2^\ast$. $\square$

\noindent\textbf{Proof of Proposition \ref{NP_thm_KKT_form}:}
    For any $\varphi\in \Phi$,
    \begin{align*}
        \int \varphi g d\nu - \sum_{i=1}^{M_0} \tilde\lambda_i \biggl( \int \varphi f_{\tilde\theta_i} d\nu - \alpha\biggr)
        = & \int \varphi \biggl( g - \sum_{i=1}^{M_0} \tilde\lambda_i f_{\tilde\theta_i} \biggr) d\nu + \alpha \sum_{i=0}^{M_0} \tilde\lambda_i,
    \end{align*}
    which is clearly maximized at $\tilde\varphi_{\widetilde\Lambda}$.  $\square$

\noindent\textbf{Proof of Theorem \ref{EMW_convergence}:}
    The proof follows from Theorem 3.2.2 of \cite{nesterov2018lecture}. It only remains to bound the Lipschitz constant of $\tilde\phi$ and derive its subdifferential, which is done in Lemma \ref{dual_subdiff}.  The second claim follows from equations (3.2.16)--(3.2.21) in section 3.2.3 in \cite{nesterov2018lecture}. $\square$

\noindent\textbf{Proof of Theorem \ref{outer_loop_convergence}:} The proof follows from Theorem 3.2.2 of \cite{nesterov2018lecture}. It only remains to bound the Lipschitz constant of $\phi$ and derive its subdifferential, which is done in Lemma \ref{outer_loop_subdiff}.  The second claim follows from equations (3.2.16)--(3.2.21) in section 3.2.3 in \cite{nesterov2018lecture}. $\square$

\subsection{Auxiliary Results}

\begin{lemma}[Based on Theorem 6.1.5 of \cite{ruschendorf2014statistik}]\label{existence_minimax_test}
    For $\mathbb{P}_\theta$ the probability measure associated with density function $f_\theta$, if $\mathcal{P} = \{\mathbb{P}_\theta : \theta\in \Theta\} \ll \nu$, there exists a test $\varphi^\ast \in \Phi_\alpha$ for any $\alpha\in [0,1]$ satisfying
    \begin{align*}
        \inf_{\theta \in \Theta_1} \int (\varphi^\ast - \varphi_{ah}) f_\theta d\nu
        = \sup_{\varphi\in \Phi_\alpha} \inf_{\theta\in \Theta_1} \int (\varphi-\varphi_{ah} ) f_\theta d\nu.
    \end{align*}
\end{lemma}
\noindent\textbf{Proof:}
    The existence result relies on properties of the set of test functions $\Phi$ and the weak-$\ast$ topology on the space of measurable and $\nu$-essentially bounded functions $\mathcal{L}_\infty(\nu) := \{\varphi \in \mathcal{L}(\mathcal{Y}) \,:\, \exists K\in \mathbb{R} \text{ s.th. } \lvert \varphi \rvert \le K \, \nu-\text{a.e.}\}$, where $\mathcal{L}(\mathcal{Y})$ denotes the set of measurable functions mapping $\mathcal{Y}$ into $\mathbb{R}$.
    As usual, we endow $\mathcal{L}_\infty(\nu)$ with the norm $\lVert \varphi \rVert _\infty := \inf\{ K: \lvert \varphi \rvert \le K \,\text{ } \nu-\text{a.e.}\}$ and treat functions $\varphi, \varphi' \in \mathcal{L}_\infty (\nu)$ as equal when $\varphi = \varphi'$ $\nu$-a.e.
    Further, a sequence $\{\varphi_n\} \subset \mathcal{L}_\infty(\nu)$ converges with respect to the weak-$\ast$ topology if
    \begin{align*}
        \int \varphi_n \delta d\nu \to \int \varphi \delta d\nu \qquad \forall \delta\in \mathcal{L}_1(\nu),
    \end{align*}
    as $n\rightarrow\infty$, where $\mathcal{L}_1(\nu)$ denotes the space of (equivalence classes) of absolutely $\nu$-integrable functions.
    
    The importance of the weak-$\ast$ topology lies first in the Banach-Alaoglu theorem, according to which the closed unit ball $B = \{ \varphi \in L_\infty(\nu) : \lVert \varphi\rVert _\infty \le 1\}$ is weak-$\ast$ compact. Since $\mathcal{P} \ll \nu$, $\Phi$ is a weak-$\ast$ closed subset of $B$ and therefore $\Phi$ is weak-$\ast$ compact.
    Secondly, the power function $\beta : \Phi \to [0,1]^{\Theta}$, $\varphi \mapsto \beta_\varphi$ is weak-$\ast$ continuous. In order to see this, note that $\beta$ is continuous if and only if $\varphi \mapsto \beta_\varphi(\theta)$ is continuous for all $\theta\in \Theta$. Now, since for all $\theta \in \Theta$, 
    \begin{align*}
        \beta_{\varphi}(\theta) = \int \varphi \frac{d\mathbb{P}_\theta}{d\nu} d\nu 
    \end{align*}
    with $\frac{d\mathbb{P}_\theta}{d\nu} \in \mathcal{L}_1(\nu)$, the continuity follows by the definition of weak-$\ast$ convergence, given above.
    Thus, not only is $\Phi$ weak-$\ast$ compact, but also the set of power function differences $G:= \{\beta_\varphi - \beta_{ah} : \Theta \to [-1,1] \,\vert \, \forall \theta \in \Theta \text{ s.th. } \beta_\varphi (\theta) - \beta_{ah}(\theta) = \int (\varphi - \varphi_{ah}) f_\theta d\nu \}$ is weak-$\ast$ compact as a continuous image of a compact set.\footnote{See Theorem 6.1.4 in \cite{ruschendorf2014statistik}.}

    These observations readily imply the claimed existence of such minimax-type tests. Since 
    \begin{align*}
        \Phi_\alpha = \bigcap_{\theta\in \Theta_0} \{ \varphi\in \Phi: \beta_\varphi (\theta) \le \alpha\},
    \end{align*}
    $\Phi_\alpha$ is a weak-$\ast$ closed subset of $\Phi$ and hence weak-$\ast$ compact.
    Now, let $\kappa: \Phi_\alpha \to [-1,1]$ be defined as $\kappa(\varphi) := \inf_{\theta\in \Theta_1} \{\beta_\varphi (\theta) - \beta_{ah}(\theta)\}$. 
    By continuity of $\varphi\mapsto \beta_\varphi (\theta) - \beta_{ah}(\theta)$, $\kappa$ is upper semicontinuous and hence attains its supremum on the compact set $\Phi_\alpha$. Thus, there exists $\varphi^\ast \in \Phi_\alpha$ such that 
    \begin{align*}
        \kappa(\varphi^\ast) = \inf_{\theta\in \Theta_1} \{\beta_{\varphi^*} (\theta) - \beta_{ah}(\theta)\}
        = \sup_{\varphi \in \Phi_\alpha} \inf_{\theta \in \Theta_1} \{\beta_{\varphi} (\theta) - \beta_{ah}(\theta)\}
    \end{align*}
    and $\varphi^\ast$ is a minimax test.
$\square$
\begin{lemma}\label{minimax_theorem}
    If $\mathcal{P} = \{\mathbb{P}_\theta : \theta\in \Theta\} \ll \nu$, then
    \begin{align*}
        \sup_{\varphi \in \Phi_\alpha} \inf_{\Omega \in \Delta} \iint (\varphi - \varphi_{ah} ) f_\theta d\nu d\Omega(\theta) 
        = \inf_{\Omega \in \Delta} \sup_{\varphi \in \Phi_\alpha} \iint (\varphi - \varphi_{ah} ) f_\theta d\nu d\Omega(\theta),
    \end{align*}
    where $\Delta$ denotes the set of probability measures over either $\Theta_1$ or $\Delta_{M_1}$.
\end{lemma}
\noindent\textbf{Proof:}
    Note that $\Delta$ and $\Phi_\alpha$ are convex and that $r(\Omega, \varphi) :=\iint (\varphi - \varphi_{ah} ) f_\theta d\nu d\Omega(\theta) $ is linear in both of its arguments. Further, by similar arguments to those in the proof of Lemma \ref{existence_minimax_test}, $\varphi \mapsto r(\Omega, \varphi)$ is weak-$\ast$ continuous for all $\Omega \in \Delta$ and $\Phi_\alpha$ is weak-$\ast$ compact.  
    We can further endow $\Delta$ with the weak topology on $\mathcal{L}_1(\nu)$ when we identify $\Omega$ with its implied density $h_\Omega = \int f_\theta d\Omega (\theta) \in \mathcal{L}_1(\nu)$.
    Hence, the Ky Fan minimax theorem\footnote{See Theorem 2 in \cite{fan1953minimax}.} applies and implies
    \begin{align*}
        \sup_{\varphi \in \Phi_\alpha} \inf_{\Omega\in\Delta} r(\Omega, \varphi) 
        = \inf_{\Omega\in\Delta} \sup_{\varphi \in \Phi_\alpha} r(\Omega, \varphi) .
    \end{align*}
    This proves the claim.
$\square$

\begin{lemma}\label{pot_power_fun_inner_loop_new}
    Suppose $\alpha\in(0,1)$ and $\Omega \in \Delta$, where $\Delta$ denotes the set of probability measures over either $\Theta_1$ or $\Delta_{M_1}$, and consider the problem
    \begin{align}
        \sup_{\varphi \in \Phi_\alpha} \int (\varphi - \varphi_{ah}) \int f_\theta d\Omega (\theta) d\nu. \label{inner_loop_non_discretized_new}
    \end{align}
    Then, the following statements hold:
    \begin{enumerate}
        \item There exists a solution to \eqref{inner_loop_non_discretized_new}, i.e., 
        \begin{align*}
            \Phi_{\Omega}^\ast := \argmax_{\varphi\in \Phi_\alpha} \int (\varphi - \varphi_{ah}) \int f_\theta d\Omega (\theta) d\nu \neq \emptyset.
        \end{align*}
        \item Let $\mathcal{M}_0$ denote the set of finite measures over $\Theta_0$. For any $\Lambda\in \mathcal{M}_0$,
        \begin{align*}
            \Phi_{\Omega, \Lambda}^\ast := &\argmax_{\varphi\in \Phi} \int (\varphi - \varphi_{ah}) \int f_\theta d\Omega (\theta) d\nu
            - \int \biggl(\int \varphi f_\theta d\nu - \alpha\biggr) d\Lambda (\theta) \neq \emptyset
        \end{align*}
        and any $\varphi \in \Phi_{\Omega, \Lambda}^\ast$ can be written as
        \begin{align*}
            \varphi = \begin{cases}
                1 & \text{, if } \int f_\theta d\Omega (\theta) > \int f_\theta d\Lambda (\theta)\\
                \varkappa & \text{, if } \int f_\theta d\Omega (\theta) = \int f_\theta d\Lambda (\theta)\\
                0 & \text{, if } \int f_\theta d\Omega (\theta) < \int f_\theta d\Lambda (\theta)
            \end{cases}
        \end{align*}
        for some $\varkappa\in \Phi$.
        \item It holds that \eqref{inner_loop_non_discretized_new} is equal to
        \begin{align*}
    \inf_{\Lambda \in \mathcal{M}_0} \sup_{\varphi\in \Phi} \int (\varphi - \varphi_{ah}) \int f_\theta d\Omega (\theta) d\nu
            - \int \biggl(\int \varphi f_\theta d\nu - \alpha \biggr)d\Lambda (\theta).
        \end{align*}
        \item Let 
        \begin{align*}
            \mathcal{M}_0^\ast = \argmin_{\Lambda \in \mathcal{M}_0} \ \sup_{\varphi\in \Phi} \int (\varphi - \varphi_{ah}) \int f_\theta d\Omega (\theta) d\nu
            - \int \biggl(\int \varphi f_\theta d\nu - \alpha\biggr) d\Lambda (\theta).
        \end{align*}
        Assume that $\mathcal{M}_0^\ast \neq \emptyset$ and that the event 
        \[
            \biggl\{y\in \mathcal{Y}: \int f_\theta d\Omega (\theta) = \int f_\theta d\Lambda^\ast (\theta)\biggr\}
        \]
        has $\nu$-measure zero for all $\Lambda^\ast\in \mathcal{M}_0^\ast$, such that for any $\Lambda^\ast\in \mathcal{M}_0^\ast$ there is a $\nu$-a.e.\ unique element in $\Phi^*_{\Omega,\Lambda^*}$ given by
        \[
           \varphi^*_{\Omega,\Lambda^*} =  \mathbbm{1}\biggl\{\int f_\theta d\Omega (\theta) > \int f_\theta d\Lambda^\ast (\theta)\biggr\}.
        \] 
        Then, for any $\Lambda^\ast_1, \Lambda^\ast_2 \in \mathcal{M}_0^\ast$ and any $\tilde \varphi_1, \tilde \varphi_2 \in \Phi_{\Omega}^\ast$
        \[
           \varphi^*_{\Omega,\Lambda^*_1} = \varphi^*_{\Omega,\Lambda^*_2} =  \tilde \varphi_1 = \tilde \varphi_2 \ \nu\text{-a.e.}
        \]
        \item If $\Lambda^\ast \in \mathcal{M}_0^\ast$ with $\Lambda^\ast(\Theta_0) > 0$, then $\tilde \Lambda^\ast = \Lambda^\ast / \Lambda^\ast(\Theta_0)$ is a least favorable distribution.
        \item If there exists a least favorable distribution $\Lambda_\Omega^\ast$, then $\mathrm{cv}_{\Omega} \Lambda_\Omega^\ast \in \mathcal{M}_0^\ast$, where $\mathrm{cv}_{\Omega}$ denotes the critical value of the Neyman-Pearson test of $\int f_\theta d\Lambda_\Omega^\ast$ against $\int f_\theta d\Omega$.
    \end{enumerate}
\end{lemma}

\noindent
\textbf{Proof:} We prove each part of the lemma as follows:
\begin{enumerate}
    \item This follows by similar arguments to those in the proof of Lemma \ref{existence_minimax_test}.

    \item Rewrite the criterion function as follows
    \begin{align*}
        &\int (\varphi - \varphi_{ah}) \int f_\theta d\Omega (\theta) d\nu
        - \int \biggl(\int \varphi f_\theta d\nu - \alpha\biggr) d\Lambda (\theta)\\
        = &\int \varphi \biggl(\int f_\theta d\Omega (\theta)-\int f_\theta d\Lambda(\theta)\biggr)  d\nu
        + \alpha \Lambda (\Theta_0) -\int \varphi_{ah} \int f_\theta d\Omega (\theta) d\nu.
    \end{align*}
    This function is maximized by any $\varphi \in \Phi$ of the claimed form.

    \item By 2.~and its proof, 
    \begin{align*}
        &\inf_{\Lambda \in \mathcal{M}_0} \sup_{\varphi\in \Phi} \int (\varphi - \varphi_{ah}) \int f_\theta d\Omega (\theta) d\nu
            - \int \biggl(\int \varphi f_\theta d\nu - \alpha \biggr)d\Lambda (\theta)\\
        = &\inf_{\Lambda \in \mathcal{M}_0} \int \biggl(\int f_\theta d\Omega (\theta)-\int f_\theta d\Lambda(\theta)\biggr)_+  d\nu
        + \alpha \Lambda (\Theta_0) -\int \varphi_{ah} \int f_\theta d\Omega (\theta) d\nu,
    \end{align*}
    where $(a)_+ = \max\{0,a\}$ for any $a\in \mathbb{R}$.
    The claim now follows along the same lines as the proof of Theorem 4 in \cite{krafftwitting1967}.

    \item For any $\Lambda^\ast\in \mathcal{M}_0^*$ and any $\tilde \varphi \in \Phi_\Omega^\ast$, we have
    \begin{align*}
        &\int (\tilde \varphi - \varphi_{ah}) \int f_\theta d\Omega (\theta) d\nu\\
        = &\sup_{\varphi\in \Phi_\alpha} \int (\varphi - \varphi_{ah}) \int f_\theta d\Omega (\theta) d\nu\\
        = &\inf_{\Lambda\in \mathcal{M}_0} \sup_{\varphi\in \Phi}\int (\varphi - \varphi_{ah}) \int f_\theta d\Omega (\theta) d\nu
        - \int \biggl(\int \varphi f_\theta d\nu - \alpha\biggr) d\Lambda (\theta)\\
        = &\sup_{\varphi\in \Phi}\int (\varphi - \varphi_{ah}) \int f_\theta d\Omega (\theta) d\nu
        - \int \biggl(\int \varphi f_\theta d\nu - \alpha\biggr) d\Lambda^\ast (\theta)\\
        \ge &\int (\tilde \varphi - \varphi_{ah}) \int f_\theta d\Omega (\theta) d\nu
        - \int \biggl(\int \varphi^\ast f_\theta d\nu - \alpha\biggr) d\Lambda^\ast (\theta)\\
        \ge &\int (\tilde \varphi - \varphi_{ah}) \int f_\theta d\Omega (\theta) d\nu,
    \end{align*}
    where we have used $\tilde \varphi \in \Phi_\Omega^\ast$ in the first equality, part 3.\ in the second equality, $\Lambda^\ast \in \mathcal{M}_0^*\neq \emptyset$ in the third equality, the definition of the supremum together with $\tilde \varphi \in \Phi_\alpha \subset \Phi$ in the first inequality and that $\tilde \varphi \in \Phi_\alpha$ in the last inequality.
    This implies that $\tilde \varphi \in \Phi_{\Omega, \Lambda^\ast}^\ast$ and thus $\tilde \varphi = \varphi^*_{\Omega,\Lambda^*} \nu$-a.e. Since $\tilde \varphi$ and $\Lambda^*$ are arbitrary, the desired result follows.

    \item[5./6.] The problem in part 3.\ leads to the same set of solutions as the problem given in equations (22) and (23) in \cite{krafftwitting1967}. The proof therefore follows along the same lines as the proof of Theorem 12 in \cite{krafftwitting1967}. $\square$
    \end{enumerate}

\begin{lemma}\label{dual_subdiff}
The following statements hold:
    \begin{enumerate}
        \item $\tilde\phi$ is Lipschitz continuous such that
        \begin{align*}
            \lvert \tilde\phi(\widetilde\Lambda) - \tilde\phi(\widetilde\Lambda') \rvert \le \sqrt{M_0 \max\{ 1-\alpha, \alpha\}} \lVert \widetilde\Lambda - \widetilde\Lambda'\rVert _2
        \end{align*}
        for any $\widetilde\Lambda, \widetilde\Lambda' \ge 0$.
        
        \item For any $\widetilde\Lambda\geq 0$,
        \begin{align*}
            \biggl(\alpha - \int \varphi_{\widetilde\Lambda} f_{\tilde\theta_i} d\nu \biggr)_{i=1}^{M_0} \in \partial \tilde\phi(\widetilde\Lambda).
        \end{align*}
    \end{enumerate}
\end{lemma}

\noindent\textbf{Proof:} We prove each part of the lemma as follows:
\begin{enumerate}
    \item Take any $\widetilde\Lambda, \widetilde\Lambda' \ge 0$. Then, by standard bounds for the supremum and the Cauchy-Schwarz inequality,
    \begin{align*}
        \lvert \tilde\phi(\widetilde\Lambda) - \tilde\phi(\widetilde\Lambda') \rvert 
        \le &\sup_{\varphi \in \Phi} \biggl\lvert \sum_{i=1}^{M_0} \{\tilde\lambda_i - \tilde\lambda_i'\} \biggl( \int \varphi f_{\tilde\theta_i} d\nu - \alpha\biggr)\biggr\rvert\\
        \le &\lVert \widetilde\Lambda - \widetilde\Lambda'\rVert _2 \sup_{\varphi\in \Phi} \sqrt{\sum_{i=1}^{M_0} \biggl( \int \varphi f_{\tilde\theta_i} d\nu - \alpha\biggr)^2}.
    \end{align*}
    The supremum on the right hand side is achieved by either $\varphi = 1$ or $\varphi = 0$ and thus
    \begin{align*}
        \sup_{\varphi\in \Phi} \sqrt{\sum_{i=1}^{M_0} \biggl( \int \varphi f_{\tilde\theta_i} d\nu - \alpha\biggr)^2} = \sqrt{M_0 \max\{ 1-\alpha, \alpha\}}.
    \end{align*}
    
    \item By definition of $\tilde\phi$ as a maximum and Proposition \ref{NP_thm_KKT_form}, for any $\widetilde\Lambda, \widetilde\Lambda' \ge 0$,
    \begin{align*}
        \tilde\phi(\widetilde\Lambda') \ge &\int \varphi_{\widetilde\Lambda} d\nu - \sum_{i=1}^{M_0} \tilde\lambda_i' \biggl( \int \varphi_{\widetilde\Lambda} f_{\tilde\theta_i} d\nu - \alpha\biggr)\\
        = &\int \varphi_{\widetilde\Lambda} d\nu - \sum_{i=1}^{M_0} \tilde\lambda_i \biggl( \int \varphi_{\widetilde\Lambda} f_{\tilde\theta_i} d\nu - \alpha\biggr) - \sum_{i=1}^{M_0} \{\tilde\lambda_i'-\tilde\lambda_i\} \biggl( \int \varphi_{\widetilde\Lambda} f_{\tilde\theta_i} d\nu - \alpha\biggr)\\
        = &\tilde\phi(\widetilde\Lambda) + \sum_{i=1}^{M_0} \{\tilde\lambda_i'-\tilde\lambda_i\} \biggl( \underbrace{\alpha - \int \varphi_{\widetilde\Lambda} f_{\tilde\theta_i} d\nu}_{=: k_i}\biggr)
    \end{align*}
    and thus $\tilde\phi(\widetilde\Lambda') \ge \tilde\phi(\widetilde\Lambda) + k^\intercal (\widetilde\Lambda' - \widetilde\Lambda)$, for all $\widetilde\Lambda' \ge 0$, i.e., $k\in \partial \tilde\phi(\widetilde\Lambda)$. $\square$
    \end{enumerate}

\begin{lemma}\label{outer_loop_subdiff}
The following statements hold:
    \begin{enumerate}
        \item $\phi$ is Lipschitz continuous such that
        \begin{align*}
            \lvert \phi(\bar\Omega) - \phi(\bar\Omega') \rvert \le \sqrt{M_1} \lVert \bar\Omega - \bar\Omega'\rVert _2
        \end{align*}
        for any $\bar\Omega, \bar\Omega' \in\Delta_{M_1}$.
        
        \item For any $\bar\Omega\in \Delta_{M_1}$,
        \begin{align*}
            \biggl(\int (\varphi_{\bar\Omega}^* - \varphi_{ah}) f_{\theta_i} d\nu \biggr)_{i=1}^{M_1} \in \partial \phi(\bar\Omega).
        \end{align*}
    \end{enumerate}
\end{lemma}


\noindent\textbf{Proof:} We prove each part of the lemma as follows:
\begin{enumerate}
    \item Take any $\bar\Omega, \bar\Omega' \in \Delta_{M_1}$. Then, by standard bounds for the supremum and the Cauchy-Schwarz inequality,
    \begin{align*}
        \lvert \phi(\bar\Omega) - \phi(\bar\Omega') \rvert 
        \le &\sup_{\varphi \in \Phi_\alpha} \biggl\lvert \sum_{i=1}^{M_1} \{\omega_i - \omega_i'\} \int (\varphi - \varphi_{ah}) f_{\theta_i} d\nu \biggr\rvert\\
        \le &\lVert \bar\Omega - \bar\Omega'\rVert _2 \sup_{\varphi\in \Phi_\alpha} \sqrt{\sum_{i=1}^{M_1} \biggl( \int (\varphi - \varphi_{ah}) f_{\theta_i} d\nu\biggr)^2}.
    \end{align*}
    The supremum on the right hand side can be bounded above using
    \begin{align*}
        \sup_{\varphi\in \Phi} \sqrt{\sum_{i=1}^{M_1} \biggl( \int (\varphi - \varphi_{ah}) f_{\theta_i} d\nu\biggr)^2} \le \sqrt{M_1}.
    \end{align*}

    \item By definition of $\phi$ as a maximum, for any $\bar\Omega, \bar\Omega' \in \Delta_{M_1}$,
    \begin{align*}
        \phi(\bar\Omega') \ge &\sum_{i=1}^{M_1} \omega_i' \int (\varphi_{\bar\Omega}^* - \varphi_{ah}) f_{\theta_i} d\nu\\
        = &\sum_{i=1}^{M_1} \omega_i \int (\varphi_{\bar\Omega}^* - \varphi_{ah}) f_{\theta_i} d\nu
        + \sum_{i=1}^{M_1} (\omega_i' - \omega_i) \underbrace{\int (\varphi_{\bar\Omega}^* - \varphi_{ah}) f_{\theta_i} d\nu}_{=:\gamma_i}\\
        = &\phi(\bar\Omega) + \gamma^\intercal (\bar\Omega' - \bar\Omega)
    \end{align*}
    and thus $\gamma\in \partial \phi(\bar\Omega)$. $\square$
    \end{enumerate}

\subsection{Details for the Applications Considered in Section \ref{sec:applications}} \label{app:details:applications}

For all applications we use 300,000 simulation draws to evaluate rejection probabilities. The choice of $\{h_k\}$ is common across all applications. In particular, 
\[
h_k = \begin{cases} 0.01 & \text{ if } \min_{i=1,\dots,M_1}\{\gamma_i\} < -0.02\\
        0.001 & \text{ if } -0.02 \leq \min_{i=1,\dots,M_1}\{\gamma_i\} < -0.002\\
        0.0001 & \text{ otherwise, }
\end{cases}
\]
where $\gamma = (\gamma_1,\dots, \gamma_{M_1})$ is defined in step 2.(a) of the algorithm in Section \ref{sec:outer loop alg}. Furthermore, the number of iterations for the outer loop is set equal to 1,000. 

\subsubsection{Details for the Homoskedastic Linear IV Model}

The draws of $Q$ are obtained through baseline draws of $S$ and $T$, given by two independent sets of draws from $\mathcal{N}(0,I_k)$, which are standardized across all simulation draws. The conditional critical values for the CLR test are obtained using 1,000,000 simulation draws. The choice of $\{\tilde h_k\}$ is common for the two designs (fixed-$\Omega$ and fixed-$\Sigma$). In particular, 
\[
\tilde h_k = \begin{cases} 0.01 & \text{ if } \max_{i=1,\dots,M_0}\{\tau_i\} > 0.02\\
        0.001 & \text{ if } 0.02 \geq \max_{i=1,\dots,M_0}\{\tau_i\} > 0.002\\
        0.0001 & \text{ otherwise, }
\end{cases}
\]
where $\tau = (\tau_1,\dots, \tau_{M_0})$ is defined in step 2.(a) of the algorithm in Section \ref{sec:inner loop alg}. Furthermore, the number of iterations for the inner loop is set equal to 1,000. In both designs, we take $\Theta_0^f = \{ (\beta,\lambda): \beta = 0, \lambda \in \{0,2,\dots,150\}\}$ and $\Theta_1^f = \{ (b/\sqrt{\lambda},\lambda) : b\in \{-3.5,-3,\dots,-0.5,0.5,1,\dots,3.5\},\ \lambda \in \{ 0.1,10,20,\dots,170 \} \}$. And in the fixed-$\Sigma$ design, we take $\bar \Theta_0 = \{ (\beta,\lambda) : \beta = 0,\ \lambda \in \{1,5,10,15,20,30,40,50,70,\dots,150,175,\dots,300\} \}$ and 
\begin{align*}
    \bar \Theta_1 =& \{ (\beta,\lambda) : \beta \in \{-40,-30,-20,-10,-2.5,-1,1,6,20,30\},\ \lambda = 1 \} \\
    \cup& \{ (b/\sqrt{\lambda},\lambda) : b\in \{-40,-30,-20,-10,-5,-1,1,5,10,20,30\},\ \lambda = 5 \}\\
    \cup& \{ (b/\sqrt{\lambda},\lambda) : b\in \{-40,-30,-20,-10,-6,-1,1,5,10,20,30\},\ \lambda = 10 \}\\
    \cup& \{ (b/\sqrt{\lambda},\lambda) : b\in \{-40,-30,-20,-10,-7.5,-2,2,10,20,30\},\ \lambda = 15 \}\\
    \cup& \{ (b/\sqrt{\lambda},\lambda) : b\in \{-30,-10,-5,-3,3,7,10,20,40\},\ \lambda = 20 \}\\
    \cup& \{ (b/\sqrt{\lambda},\lambda) : b\in \{-3,-1,2,4,6,8\},\ \lambda = 30 \}\\
    \cup& \{ (b/\sqrt{\lambda},\lambda) : b\in \{-3,2,4,6,8\},\ \lambda = 40 \}\\
    \cup& \{ (b/\sqrt{\lambda},\lambda) : b\in \{-3,2,4\},\ \lambda \in \{ 50,70,\dots,150,175,\dots,300 \} \}.
\end{align*}



\begin{figure}[h]
  \centering
  \begin{subfigure}[b]{0.495\textwidth}
    \centering
    \includegraphics[width=\textwidth]{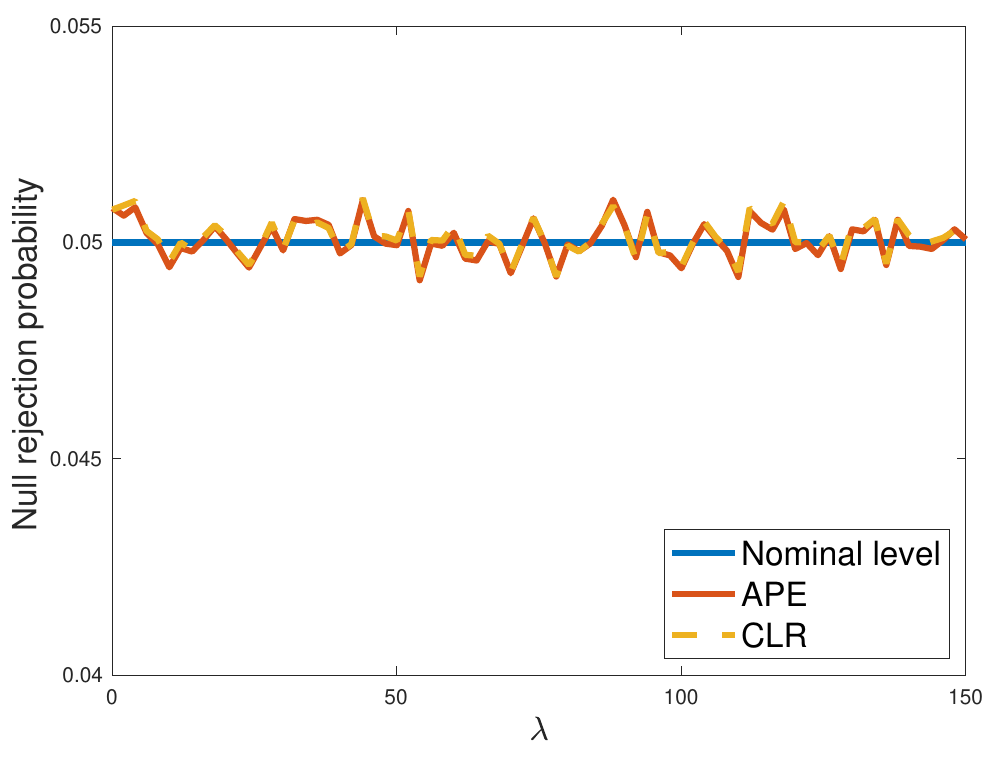}
    \caption{$k = 5$ and $\Omega_{12} = 0.5$}
  \end{subfigure}
  \hfill
  \begin{subfigure}[b]{0.495\textwidth}
    \centering
    \includegraphics[width=\textwidth]{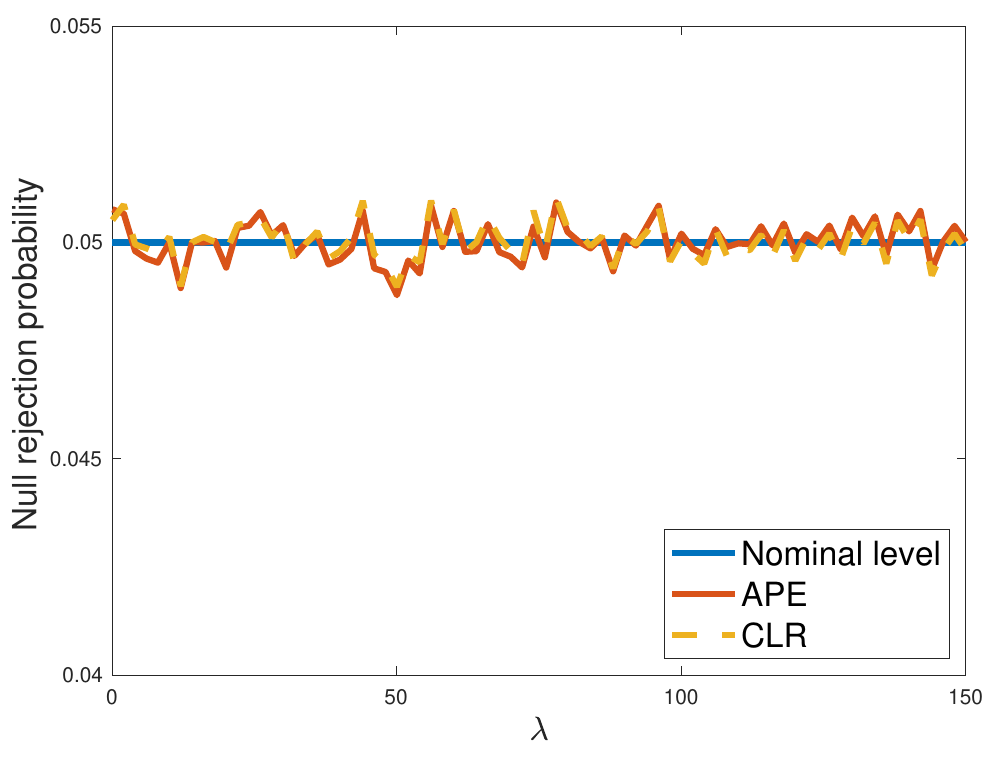}
    \caption{$k = 10$ and $\Sigma_{12} = 0.5$}
  \end{subfigure}
  \caption{Null rejection probabilities of tests underlying APEs and CLR test.}
  \label{Figure:CLR:nrp}
\end{figure}

Figure \ref{Figure:CLR:nrp} shows the null rejection probabilities of the tests underlying the APEs and of the CLR test on $\Theta_0^f$. Panel~(a) reports the results for $k = 5$ and $\Omega_{12} = 0.5$ and panel~(b) for $k = 10$ and $\Sigma_{12} = 0.5$. All probabilities are computed using 300{,}000 simulation draws, which are independent across the values in $\Theta_0^f$. In each panel, the test underlying the corresponding APE and the CLR test yield virtually identical null rejection probabilities. Although the rejection probabilities of both tests can exceed the nominal level, we know that for the CLR test any such excess is solely due to simulation error since the CLR test controls size by construction. Because the two tests coincide so closely in both designs, we conclude that the tests underlying the APEs effectively control size.

\subsubsection{Details for the Boundary Problem}
The baseline draws used to obtain the draws of $Y$ are standardized across all simulations and also symmetrized. The parameter $\tilde h_k$ is set equal to 0.01 for all $k$ and the number of iterations for the inner loop is set equal to 1,000. As mentioned in the main text, we follow \cite{EMW15} in discretizing $\Theta_0$ in terms of ``base'' distributions. Our ``base'' distributions are uniform distributions for $\delta$ on the following intervals: $[0,0.00001]$, $[0,0.04]$, $[1.99,2.01]$, $[0,0.5]$, $[0.5,1]$, \dots, $[12,12.5]$. We take $\Theta_0^f = \{ (\beta,\delta): \beta = 0, \delta \in \{0,0.1,\dots,7\}\}$ and $\Theta_1^f = \{ (\beta,\delta) : \beta \in \{-3.5,-3,\dots,-0.5,0.5,1,\dots,3.5\},\ \delta \in \{ 0,0.5,\dots,8 \} \}$.

\begin{figure}[h]
    \centering
    \includegraphics[width=0.6\textwidth]{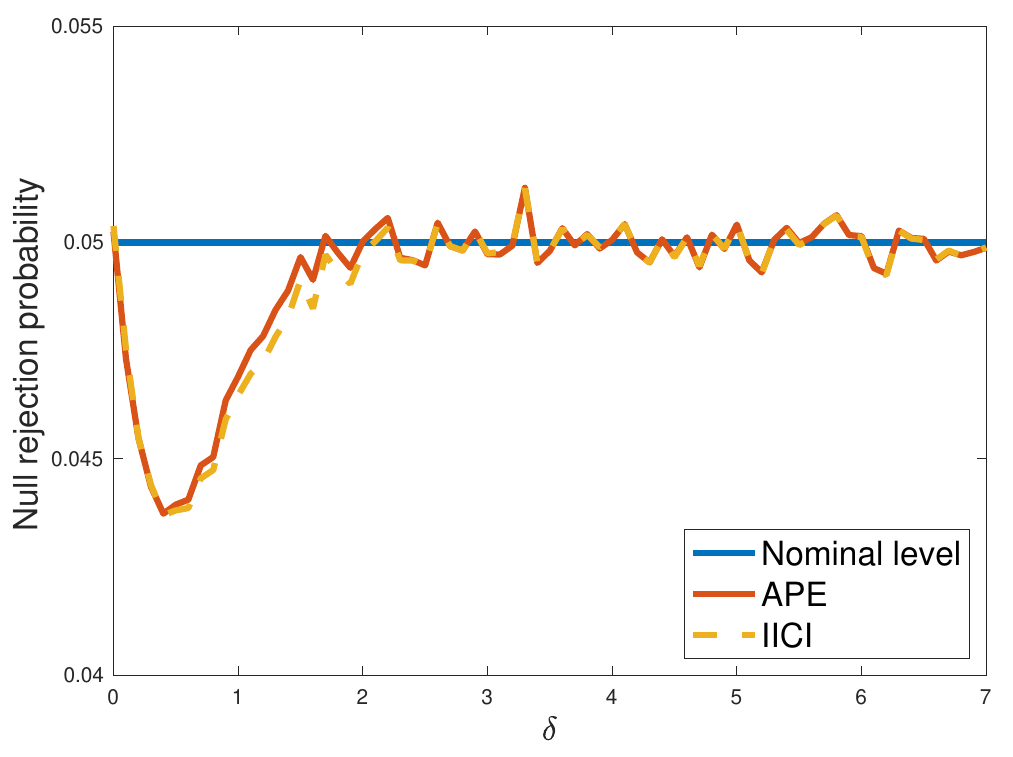}
  \caption{Null rejection probabilities of test underlying APE and test implied by IICI.}
  \label{Figure:IICI:nrp}
\end{figure}

Figure \ref{Figure:IICI:nrp} shows the null rejection probabilities of the test underlying the APE and the test implied by the IICI on $\Theta_0^f$. All probabilities are computed using 300{,}000 simulation draws, which are independent across the values in $\Theta_0^f$. Although the null rejection probabilities for both tests can exceed the nominal level, we know that for the test implied by the IICI this is only due to simulation error given that \cite{Cox24} proves that the IICI has uniformly correct coverage. Since the maximal null rejection probability of the test underlying the APE is very close to $\alpha$ and the null rejection probability of the test implied by the IICI over the entire grid, we conclude that the test underlying the APE effectively controls size.


\begin{figure}[h]
    \centering
    \includegraphics[width=0.6\textwidth]{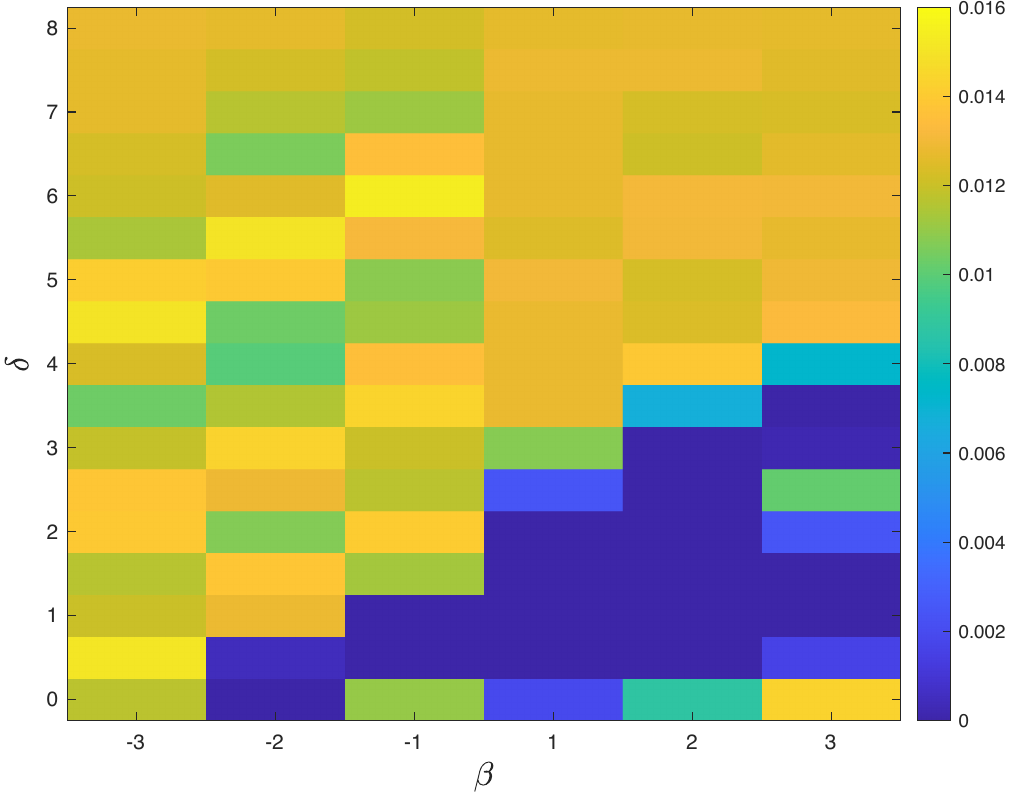}
  \caption{Weights underlying the APE for the test implied by the IICI}
  \label{Figure:IICI:weights}
\end{figure}

Figure \ref{Figure:IICI:weights} shows the weights underlying the APE corresponding to the test implied by the IICI.  Apparently the test implied by the IICI does not prioritize power at positive alternatives $\beta>0$ when the value of the nuisance parameter $\delta$ is small and $\rho=0.7$.  Interestingly, the  region of the alternative parameter space receiving little weight by the APE also roughly corresponds to where we see the largest power differences between the APE and the power of the test implied by the IICI in Figure \ref{Figure:Cox:heatmap}.

\newpage

\bibliographystyle{apalike}
\bibliography{references}


\end{document}